\documentclass[12pt,bibnotes]{iopart}
\usepackage{graphicx}
\usepackage{multirow}
\usepackage{amssymb}
\usepackage{xr}
\usepackage{filecontents} 

\def\mos2{MoS$_2$}

\begin{document}

\title[Schottky barriers, emission regimes and $R_c$ in 2H-1T' \mos2 junctions]
      {Schottky barriers, emission regimes and contact resistances in 2H-1T' \mos2 lateral metal-semiconductor junctions from first-principles}

\author{M. Laura Urquiza}
\address{Departament d'Enginyeria Electr\`onica, Universitat Aut\`onoma de Barcelona, 08193 Bellaterra, Spain}
\ead{Laura.Urquiza@uab.es}

\author{Xavier Cartoix\`a}
\address{Departament d'Enginyeria Electr\`onica, Universitat Aut\`onoma de Barcelona, 08193 Bellaterra, Spain}
\ead{Xavier.Cartoixa@uab.es}

\begin{abstract}
We have studied the finite bias transport properties of a 2H-1T' \mos2 lateral metal-semiconductor (M-S) junction by non-equilibrium Green's functions calculations, aimed at contacting the 2D channel in a field effect transistor. Our results indicate that (a) despite the fundamentally different electrostatics of line and planar dipoles, the Schottky barrier heights respond similarly to changes in doping and applied bias in 2D and 3D M-S junctions, (b) 2H-1T' \mos2 lateral junctions are free from Fermi level pinning, (c) armchair interfaces have superior contacting properties vs.\ zigzag interfaces, (d) 1T' contacts to $p$ channels will present a reduced contact resistance by a factor of 4-10 with respect to $n$ channels and (e) contacts to intermediately doped $n$ ($p$) channels operate in the field (thermionic) emission regime. We also provide an improved procedure to experimentally determine the emission regime in 2D material junctions.
\end{abstract}

\vspace{2pc}


\maketitle

\section{Introduction}

Ultrathin transition metal dichalcogenides (TMDs) have emerged as promising semiconductors to overcome the short channel effects that arise with the miniaturization of field effect transistors (FETs)~\cite{Radisavljevic2011}. 
Due to their 2D geometry and wide bandgap (in the range of 1-2 eV), these materials would reduce the direct source-drain tunnelling current and could improve the transport properties of the channel~\cite{review_elec_prop}. 
Moreover, the absence of dangling bonds outside the 2D plane gives a perfect interface with the gate and the substrate, which, together with the atomically thin structure, allows an excellent electrostatic gate control.
TMDs are also very promising for the new generation of flexible electronics; in fact, flexible and transparent FETs have been already demonstrated using semiconducting \mos2 channel~\cite{flexible_FET}. 
But TMDs in electronics are not restricted to single devices, as complex circuitry such as a \mos2 microprocessors has also been reported~\cite{mos2-cpu}. 

Despite the novel properties of these materials, the performance of TMD-based FETs is normally limited by the formation of Schottky barriers at the interfaces between the 2D channel and the metallic electrodes, which translates into a high contact resistance ($R_c$), in the range of $10^{4}$ to $10^{6}\ {\rm \Omega \! \cdot \! \mu m}$~\cite{allain}. 
Some of the attempts to reduce the negative effects of the Schottky barrier include the use of metals with low work function~\cite{sc_contact}, metals with high chalcogenide (S, Se, Te) affinity, able to provide a strong hybridization 
between the channel and the electrode~\cite{ti_contacts, dft_contact}, and the use of substitutional doping~\cite{cl-dop}. 

So far, the most promising solution for the contact resistance issue seems to be the use of phase engineering to build contacts between the semiconducting (2H) and the metallic phase of TMDs~\cite{review_phase_eng}. 
This metallic phase can be either the 1T, where one of the chalcogenide planes is rotated 60 degrees and the structure acquires inversion symmetry; or the 1T', which is a distortion/structural relaxation of the 1T phase with a lower 
energy~\cite{calandra, stability-1T, topological-1T}.
Experimentally, the presence of the 1T' phase has been observed by Eda {\it et al.}~\cite{coherent-atomic-heterostructure} and Lin {\it et al.}~\cite{atomic-mechanism}. 
The 1T' phase is used to connect the 2H channel to the metallic pads, in the same way that p$^{+}$ or n$^{-}$ doped regions are implemented in the traditional silicon-based FETs to connect the Si channel to the metallic electrodes. 
With this strategy, it was possible to achieve record low $R_c$ values of 200-300$\ {\rm \Omega \! \cdot \! \mu m}$ in \mos2-based FETs~\cite{kappera}. 
These FETs also demonstrated good performance, with mobility values of 50~cm$^{2}$/V$\cdot$s, subthreshold swing values of 90-100~mV and on/off ratios greater than $10^{7}$~\cite{kappera}.

There have been prior theoretical studies to get insight into the structural features that affect the transport properties and the contact resistance in \mos2 2H-1T interfaces~\cite{paz_palacios,Houssa2019} and 2H-1T' interfaces~\cite{saha,LiuLiShi2018}. 
However, in these works the coupled effects of the electrostatic gating and source-drain bias have not been considered, which is crucial for a complete understanding of the atomistic properties that affect the contact 
resistance under operating conditions.

In this work we investigate the electrostatic gating on the semiconducting phase of \mos2 under finite source-drain bias to understand its influence on the contact resistance of 2H-1T' \mos2 interfaces. 
We have chosen the 1T' polytype for the metallic phase because, as noted above, it has been both observed experimentally and predicted to have a lower energy than the 1T phase. 
Using density functional theory (DFT) with nonequilibrium Green's Functions (NEGF) and a physical model to emulate the electrostatic doping induced by the gate, we performed transport calculations at finite voltage to obtain I--V curves considering the effect of the gate bias on the 2H phase, and determine the regime (i.e. thermionic, tunneling, thermally assisted tunneling) under which the resulting Schottky barriers operate.

\section{Methodology}
\subsection{System description and computational details\label{ssec:system}}

The electronic structure calculations were performed within the DFT formalism as implemented in the {\sc Siesta} code~\cite{siesta}.
The exchange correlation functional was included in the Generalized Gradient Approximation (GGA) using the Perdew-Burke-Ernzerhof (PBE) parametrization~\cite{pbe}.
Although for 2D materials the GGA gaps often match experimental optical gaps despite the well known DFT underestimation of band gaps, this is because the experimental gaps are subject to strong excitonic effects that reduce the single particle gaps~\cite{Ugeda2014}. 
Since band alignments and carrier injection are expected to depend on the single particle band structure, the limitation of plain LDA/GGA still stands. This might affect the overall value of the Schottky barrier heights (SBHs) we provide in this work, but relative differences in SBHs should be more accurate~\cite{DasBlochlGunnarsson1989}.

We use norm-conserving pseudopotentials~\cite{Troullier-Martins} to describe the core electrons and a double-$\zeta$ polarized basis set for the valence electrons. 
The Brillouin zone was sampled using a grid of 9x16x1 {\bf k}-points for the 6-atom orthogonal unit cell of the 2H phase, and its equivalent for supercell calculations. 
The real space grid cutoff was set to 250~Ry.

Ionic positions were relaxed using a force tolerance of 0.03~eV/\AA\ for the {\sc Siesta} calculation and 0.05~eV/\AA\ for {\sc TranSIESTA} calculations. 
In the case of structures with interfaces, we also performed a cell optimization until the strains were lower than 0.1~GPa (1~kbar). The system studied consisted of a single layer \mos2 heterostructure, where the semiconducting phase of \mos2 (2H) is contacted laterally to the 1T' metallic phase. 
We considered two different orientations for the 2H-1T' interface, armchair (ac) and zigzag (zz). 
Finally, in order to avoid interactions between images in the non-periodic direction, a vacuum of 20~\AA\ was included.

\subsubsection{Transport calculations\label{sssec:transport}}

\begin{figure}[t!]
	\centering
	\includegraphics[width=1.0\linewidth]{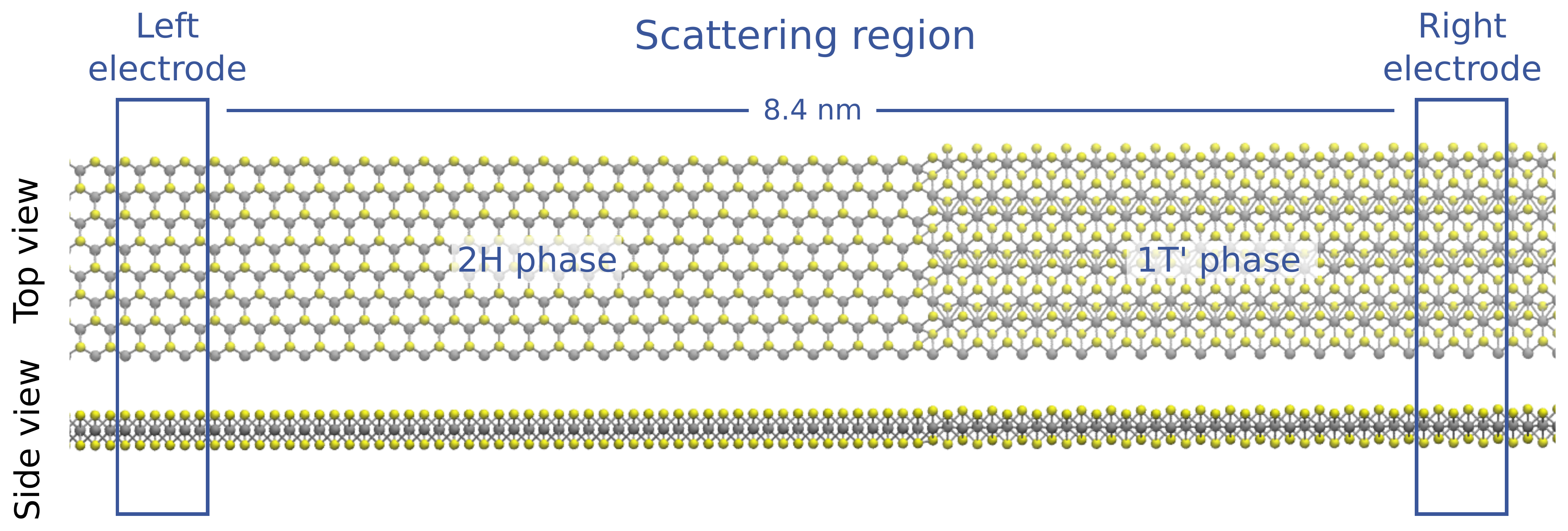}
	\caption{Device set-up for transport calculations of armchair 2H-1T' \mos2 interfaces. 
	Yellow and grey spheres represent sulphur and molybdenum atoms, respectively.}
	\label{fig:geometry}
\end{figure}

Electronic transport through the 2H-1T' interface was addressed using the NEGF formalism, as implemented in the {\sc TranSIESTA} and {\sc TBtrans} codes~\cite{transiesta}. 
To perform electronic transport with NEGF, the system must be separated into three regions: left electrode, scattering region, and right electrode, as shown in \Fref{fig:geometry}. The 2H and 1T' electrodes are extended in the semi-infinite direction to impose the open boundary conditions. 
The scattering region, which includes the interface between the 2H and 1T' phases, is 13~nm (16.5~nm) long for the armchair (zigzag) device. 
Since our system has dissimilar electrodes, we added buffer regions of pristine 2H and 1T' structures after each electrode to provide a bulk-like environment to the electrode atoms.

The transmission function is calculated as:
\begin{eqnarray}
	T_{{\bf k}_{\parallel}}(E) = \Tr \left[ G_{{\bf k}_{\parallel}} \! (E) \, \Gamma_{L,{\bf k}_{\parallel}} \! (E) \, 
	  G_{{\bf k}_{\parallel}}^\dag \! (E) \, \Gamma_{R,{\bf k}_{\parallel}} \! (E) \right] ,
	\label{eq:1}
\end{eqnarray}
where $G_{{\bf k}_{\parallel}} \! (E)$/$G_{{\bf k}_{\parallel}}^\dag \! (E)$ is the retarded/advanced Green's function, and $\Gamma_{{\bf k}_{\parallel}}$ are the 
electrode broadening matrices:
\begin{eqnarray}
	G_{{\bf k}_{\parallel}} \! (E) = [(E+i\eta) S_{{\bf k}_{\parallel}} - H_{{\bf k}_{\parallel}} - \Sigma_{L,{\bf k}_{\parallel}} (E) - \Sigma_{R,{\bf k}_{\parallel}}(E)]^{-1}\\
	\Gamma_{L/R,{\bf k}_{\parallel}}(E) = i[\Sigma_{L/R,{\bf k}_{\parallel}} (E) - \Sigma^{\dag}_{L/R,{\bf k}_{\parallel}}(E)] \ .
	\label{eq:2}
\end{eqnarray}
Here $H_{{\bf k}_{\parallel}}$, $ S_{{\bf k}_{\parallel}}$ and $\Sigma_{L/R,{\bf k}_{\parallel}}$ are the Hamiltonian, the overlap and the electrode self-energies, respectively.

The electrical current is calculated from the transmission coefficients $T_{{\bf k}_{\parallel}}(E)$ according to:
\begin{eqnarray}
	I_{L \rightarrow R} = \frac{G_{0}}{2|e|} \int \! dE \int_{BZ} \! d{\bf k}_{\parallel} \, T_{{\bf k}_{\parallel}}(E) \left[ f_{R}(E) - f_{L}(E) \right] \ ,
	\label{eq:3}
\end{eqnarray}
where $f_{R}(E)$ and $f_{L}(E)$ corresponds to the Fermi distribution 
in the right and left electrodes, $G_0$ is the quantum of conductance for the spin-degenerate case ($2 e^2 / h$), and $BZ$  denotes the integral over the Brillouin zone average~\cite{transiesta}.

\subsubsection{Electrostatic doping\label{sssec:doping}}

We simulate the charge density induced by a gate voltage ($V_{g}$)
by adding a specific fixed 
charge in the 2H region, with the total number of electrons chosen
to satisfy global charge neutrality. The electrons in the system will
self-consistenly respond to the dopant-like fixed charges when
Poisson's equation is solved. A similar strategy was performed to 
study the effect of gating on graphene nanojunctions through DFT 
calculations~\cite{nick}.
In this work, we considered channel charge densities of 
5$\times$10$^{12}$~cm$^{-2}$ and 5$\times$10$^{13}$~cm$^{-2}$. We refer 
to $p$-doped ($n$-doped) system when the mobile charge added into the channel 
is positive (negative). 

\begin{figure}[t!]
	\centering
	\includegraphics[width=0.8\linewidth]{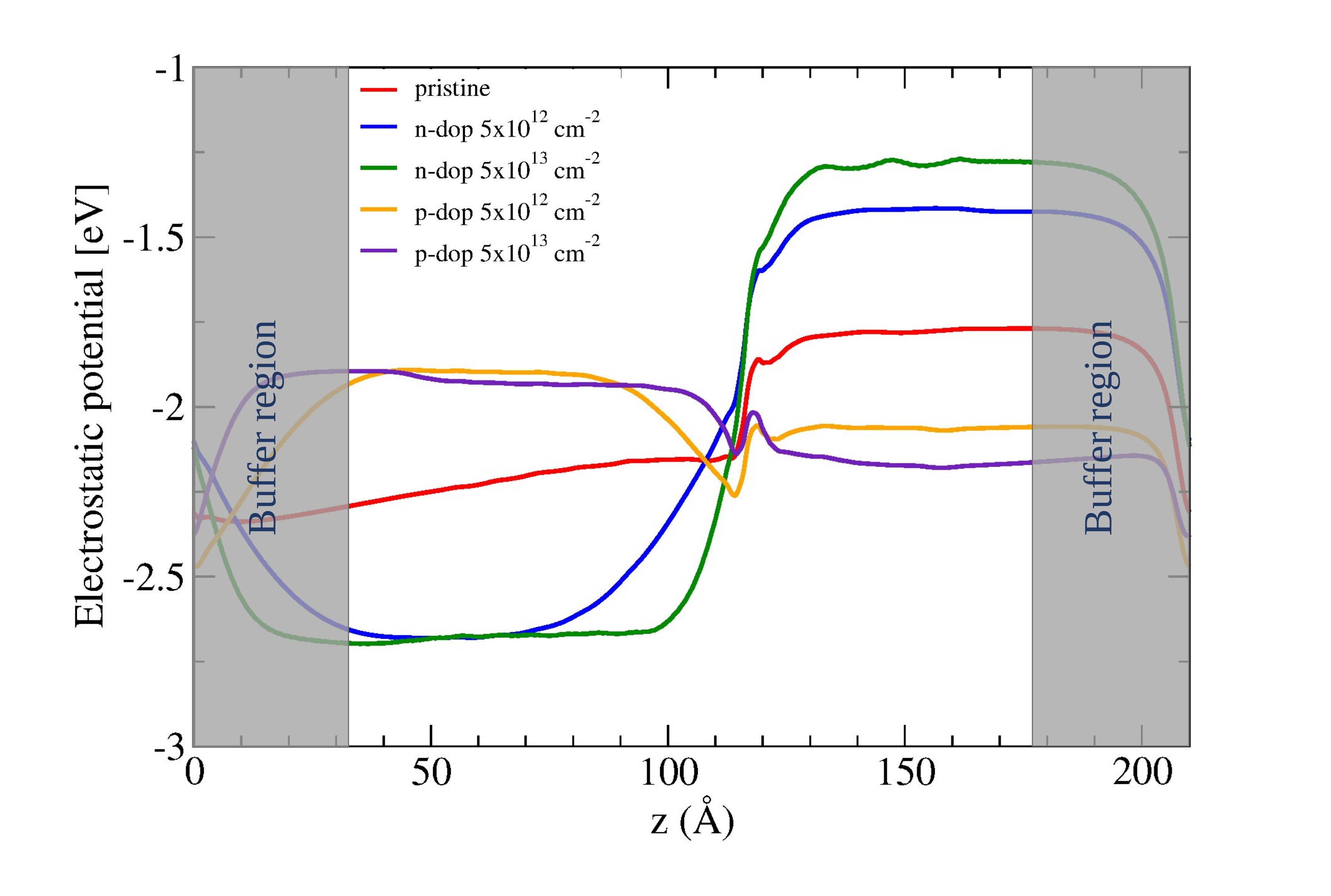}
	\caption{Electrostatic potential along the transport coordinate (z) for 
		armchair interface with different doping concentration in the 
		semiconducting phase.}
	\label{fig:macroaverage}
\end{figure}

For an accurate description of the electrostatics with NEGF, it 
is required that the electrodes behave as bulk, meaning that the 
electrostatic potential should have reached its bulk behaviour at 
the boundary between the scattering region and the semi-infinite 
bulk electrodes. Therefore, we plotted the potential profile along 
the transport direction for devices with different doping 
concentration, as shown in \Fref{fig:macroaverage}, and we noticed 
that the dipole formed at the interface is not screened in the 2H 
phase for the device without doping. As a consequence, we did not 
carry out transport calculations for this case.

\subsection{Contact resistance extraction\label{ssec:resistance}}

\begin{figure}[t!]
	\centering
	\includegraphics[width=1.0\linewidth]{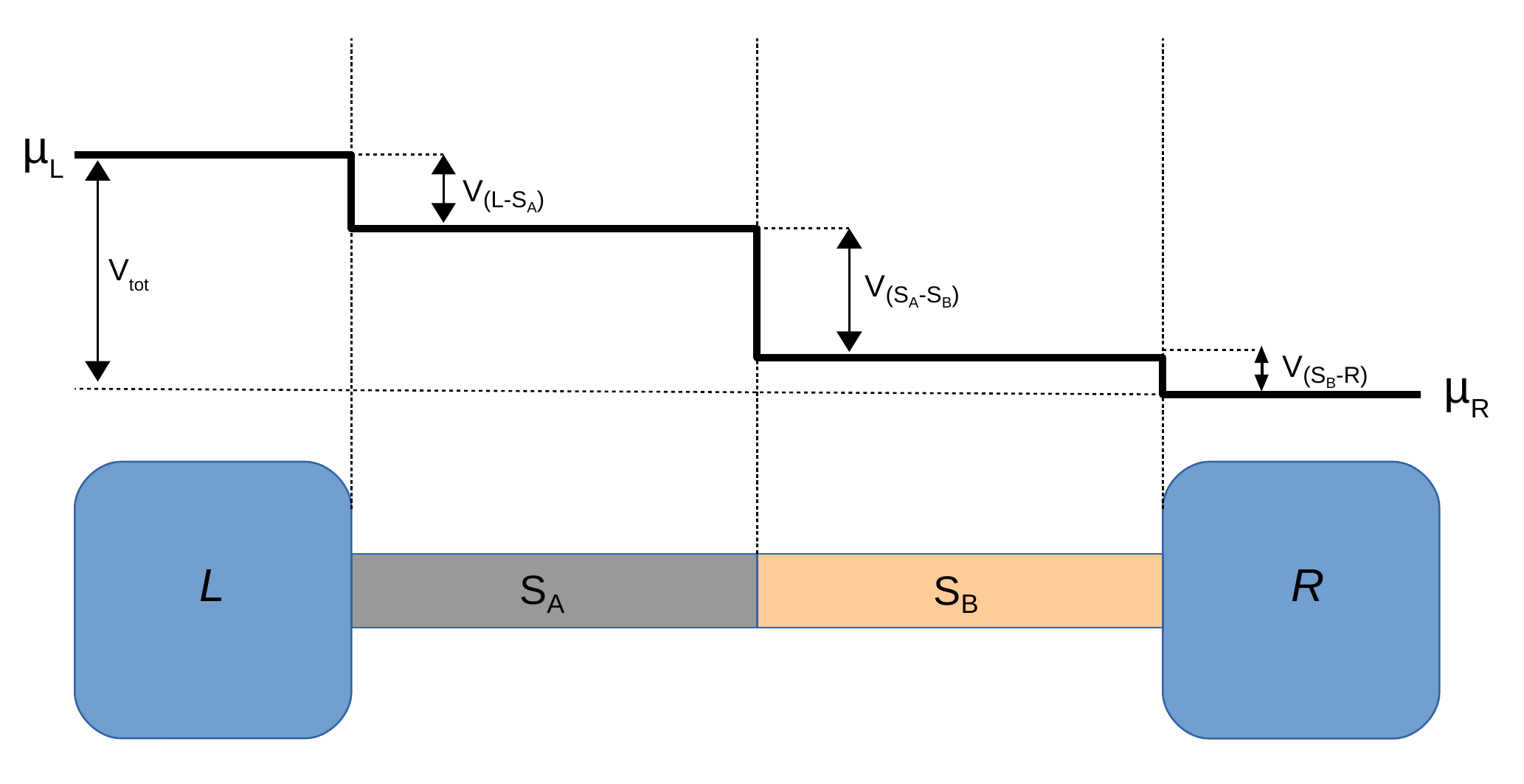}
	\caption{Voltage drop scheme for the device set-up. Here, 
		L and R represent the left and right contacts, and S$_{i}$ 
		represents the different phases present in the scattering region.} 
	\label{fig:rc-ext}
\end{figure}

The large signal contact resistance of an interface can be obtained 
from the ratio between the voltage drop across the junction and the 
current flowing through it. Using NEGF it is possible to calculate 
the current that flows through a device when a potential is applied 
at the semi-infinte contacts. In these calculations, however, the applied 
voltage ($V_{tot}$) is distributed between the interface present in 
the scattering region, ($V_{S_{A}-S_{B}}$), and the interface between 
the semi-infinite contacts with the device ($V_{L-S_{A}}$ and $V_{S_{B}-R}$), 
as seen in \Fref{fig:rc-ext}. 
Therefore, the total applied voltage can be expressed as:  
\begin{eqnarray}
	V_{tot} = V_{L-S_{A}} + V_{S_{A}-S_{B}} + V_{S_{B}-R} ,
	\label{eq:v_sum}
\end{eqnarray}
from where
\begin{eqnarray}
	R_{tot} = R_{L-S_{A}} \ + \ R_{S_{A}-S_{B}} \ + \ R_{S_{B}-R} .
	\label{eq:r_sum}
\end{eqnarray}

Here $R_{tot}$ represents the total series resistance, and can 
be obtained from the I--V curves of the whole device. 
On the other hand, $R_{L-S_{A}}$ and $R_{S_{B}-R}$ are the resistances 
between the contacts and the electrodes, which are obtained from I--V 
reference curves of pristine devices, all of the same phase. 
Since pristine devices include two electrode-contact junctions, the 
value extracted for each reference calculation must be divided by two
to consider the contribution of a simple electrode-contact interface:
\begin{eqnarray}
	R_{L-S_{A}} =  \frac{1}{2} \ \frac{V_{S_{A}-S_{A}}}{I} 
	\ \ \ {\rm and} \ \ \ 
	R_{S_{B}-R} =  \frac{1}{2} \ \frac{V_{S_{B}-S_{B}}}{I} ,
	\label{eq:rc_ref}
\end{eqnarray} 
where $V_{S_{A}-S_{A}}$ and $V_{S_{B}-S_{B}}$ are the potentials 
applied in the pristine devices to obtain the same current value
as in the whole device.

Finally, $R_{S_{A}-S_{B}}$, which is the resistance corresponding to the 
junction between the two phases, can be calculated from the other terms 
by replacing eq. (\ref{eq:rc_ref}) into eq. (\ref{eq:r_sum}) and reordering:  
\begin{eqnarray}
	R_{S_{A}-S_{B}} = \frac{V_{tot}}{I} \ - \ 
	\frac{1}{2} \ \frac{V_{S_{A}-S_{A}}}{I} \ - \ 
	\frac{1}{2} \ \frac{V_{S_{B}-S_{B}}}{I} .
	\label{eq:rc_large}
\end{eqnarray}

A similar analysis can be performed to obtain the small signal 
contact resistance, but in this case, instead of taking the ratio 
between voltage and current at each point, we take the derivative 
of the voltage with respect to the current to compute the resistance. 
Carrying this process out we have
\begin{eqnarray}
	R_{S_{A}-S_{B}}^{small} \ \Bigg|_{I_{0}} = \
	\frac{\partial V_{tot}}{\partial I} \ \Bigg|_{I_{0}} - \ 
	\frac{1}{2} \ \frac{\partial V_{S_{A}-S_{A}}}{\partial I} \ \Bigg|_{I_{0}} - \ 
	\frac{1}{2} \ \frac{\partial V_{S_{B}-S_{B}}}{\partial I} \ \Bigg|_{I_{0}} ,
	\label{eq:rc_small}
\end{eqnarray}
where derivatives are evaluated at the bias that provides a current $I_{0}$,
corresponding to the bias $V_{0}$ of interest for the whole device.

\section{Results \label{sec:results}}

\subsection{Transport regime of 2H-1T' junctions\label{ssec:transport}}

\begin{figure}[t!]
	\centering
	\includegraphics[width=1.0\linewidth]{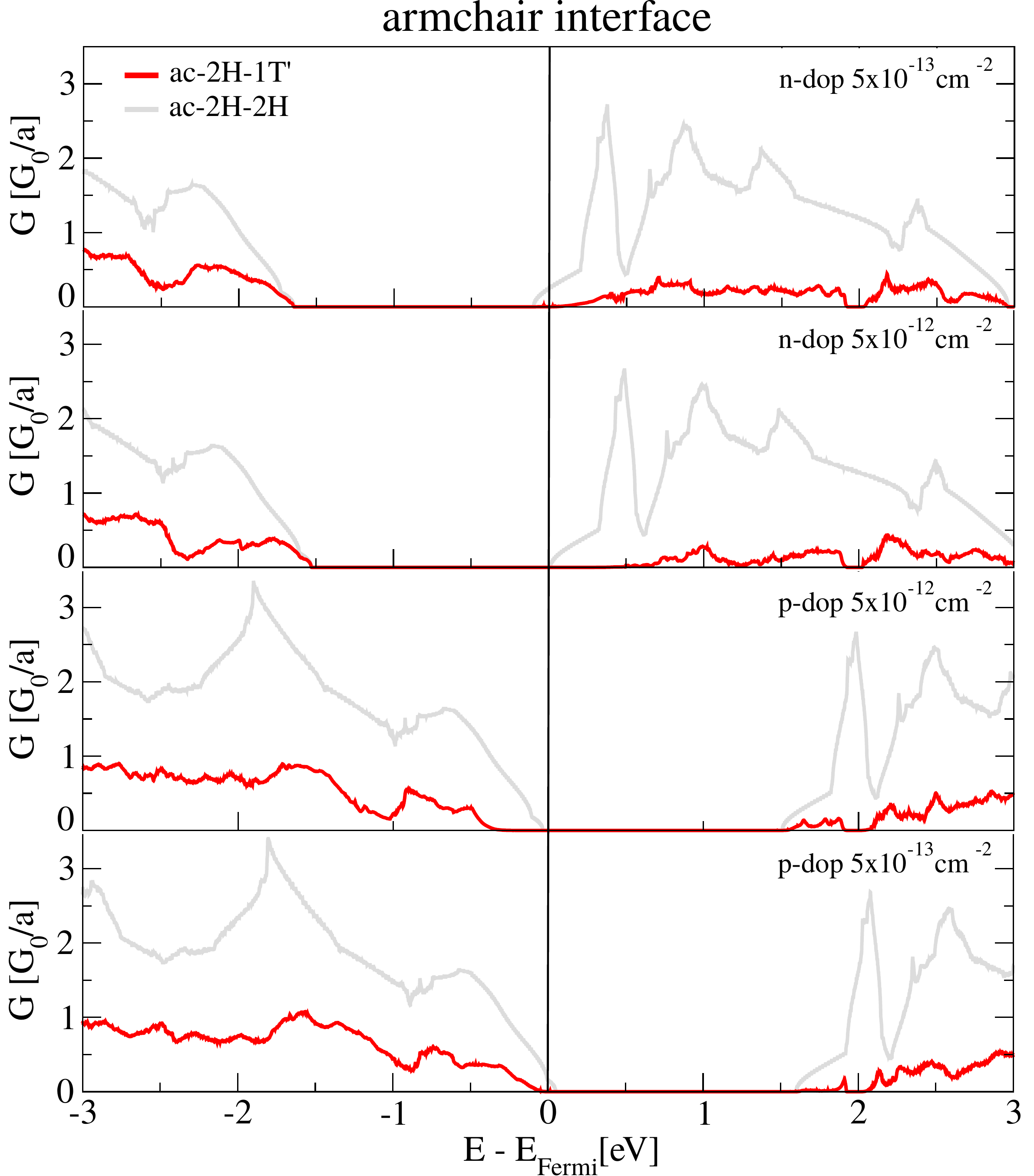}
	\caption{Specific conductance, in units of $G_{0}$ over the transverse lattice parameter  for armchair 2H-1T' and 2H-2H \mos2 interfaces with different doping concentration at zero bias.}
	\label{fig:avtrans-all}
\end{figure}

To understand the transport properties of 2H-1T' junctions, we calculated 
the specific conductance (i.e. conductance per unit of channel width) using the Landauer formula~\cite{landauer}:
\begin{eqnarray}
	G(E) a_t = \frac{2e^2}{h} \ T(E) = G_{0} \ T(E) \ , 
	\label{eq:9}
\end{eqnarray}
where $T(E)$ are the transmission coefficients averaged over all the ${\bf k}_{\parallel}$, obtained from the NEGF calculations, and $a_t$ is the width of the computational cell along the direction perpendicular to transport.

In the supplementary information (SI) we performed a detailed analysis of the transport properties across zigzag and armchair interfaces. As the results predict better conductance for armchair structures, due to the asymmetric transport in the 1T' phase, we focus our studies on ac structures. 

In \Fref{fig:avtrans-all} we show the specific conductance as a function of the energy of the incoming carrier, for armchair 2H-1T' and 2H-2H interfaces with positive and negative doping at different concentrations: intermediate (5$\times$10$^{12}$~cm$^{-2}$) and high (5$\times$10$^{13}$~cm$^{-2}$).
Comparing the transmission for the 2H-1T' device to the 2H'-2H' reference device, which provides the maximum attainable transmission, a notable reduction of the conductance is observed, mainly for the structures with intermediate doping concentration. The results also show that the injection of holes is slightly favored due to the higher number of transmission channels in the valence band.

\begin{figure}[t!]
	\centering
	\includegraphics[width=1.0\linewidth]{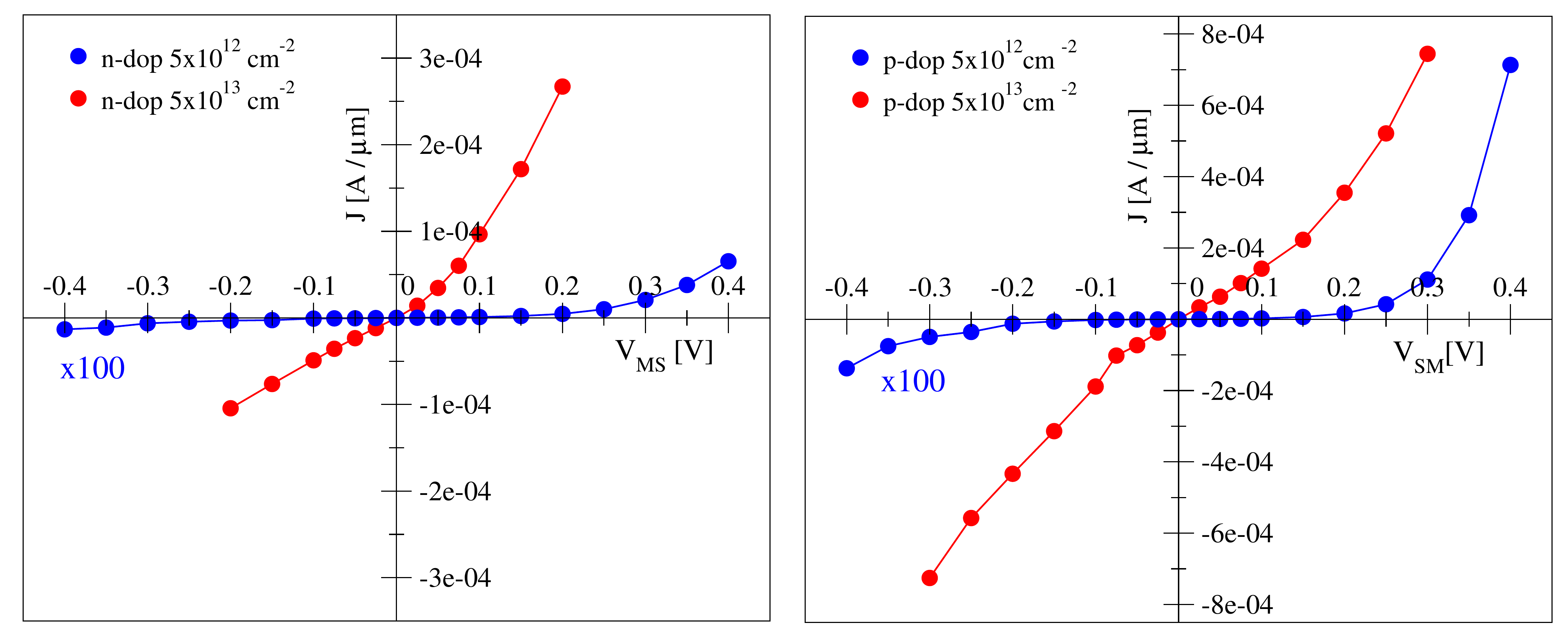}
	\caption{ Current vs. voltage curve of $n$-doped (left) and $p$-doped (right) 2H-1T'
		armchair structures.}
	\label{fig:iv}
\end{figure}

We also performed NEGF calculations at finite bias, showing the resulting current vs. voltage (I--V) characteristics in \Fref{fig:iv}.
For the high doping structures, we observe a slightly asymmetry at forward and reverse biases, specially in the $n$-doped structure, but overall there is an ohmic behavior in the I--V curve.
For intermediate doping concentrations, we observe an exponential increase of the current at forward bias, indicative of a Schottky regime, and a poor rectifying behavior, also typical of Schottky contacts.
In this regime, the transport mechanism can be (a) thermionic emission (TE) over the Schottky barrier (SB), (b) field emission (FE) with electrons around the Fermi level tunneling through the SB, or (c) thermionic-field emission (TFE), where the tunneling electrons contributing to the current are quite above the semiconductor Fermi level, but still below the top of the SB~\cite{PadovaniStratton1966}.

\begin{figure}[t!]
	\centering
	\includegraphics[width=1.0\linewidth]{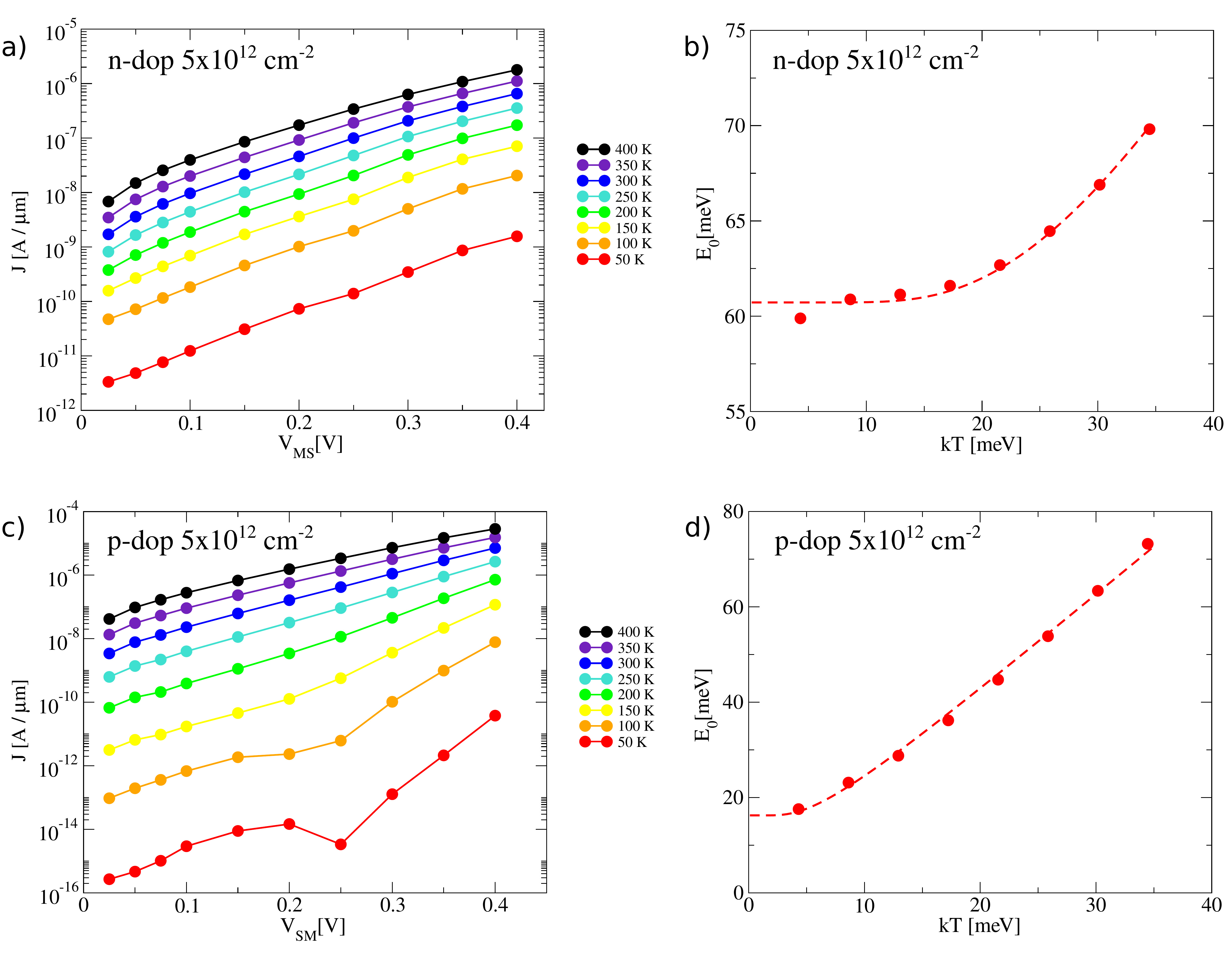}
	\caption{Forward I--V curves of a) $n$-doped and c) $p$-doped structures at different electronic temperatures, and $E_{0}$ values as a function of the temperature for b) $n$-doped and d) $p$-doped structures.}
	\label{fig:iv-temp-all}
\end{figure}

In order to determine which of these mechanisms dominates in 2H-1T' junctions, we performed a temperature study of the forward I--V characteristics. The results are shown in \Fref{fig:iv-temp-all}. Plots a) and c) show the forward bias I--V characteristic for n and p-doped structures, respectively, at different temperatures, while plots b) and d) show the energy $E_{0}$ extracted from a fit of the I--V curves to $J \propto \exp (qV/E_{0})$. 
When FE or TFE dominate, we have $E_{0}=E_{00} \coth(E_{00}/kT)$~\cite{PadovaniStratton1966}, where $E_{00}$ is an energy related to how fast the transmission coefficient through the barrier increases as the forward bias is increased. If $kT \ll E_{00}$, then FE will dominate, while when $kT \sim E_{00}$ TFE is the main mechanism~\cite{Sze2007}. On the other hand, if $kT \gg E_{00}$ then TE dominates, and we will have $E_{0} \sim \eta kT$, where $\eta$ is an ideality factor accounting for the variation of the SBH with the applied bias.

All these considerations can be summarized into the following prescription, which can also be applied to experimental measurements, allowing the determination of the dominating transport mechanism: for a range of temperatures $T$, fit the I--V curves to $J \propto \exp [qV/E_{0} (T)]$; then fit $E_{0} (T)$ to
\begin{equation}
 E_{0} (T) = E_{00} \coth (E_{00}/ \eta kT)
 \label{eq:E0vsT}
\end{equation}
and finally compare $E_{00}$ to $kT$ to determine whether FE, TFE or TE dominates, obtaining the ideality factor $\eta$ as a by-product. Performing this analysis prior to an activation-energy study should be required in order to ensure that TE dominates transport across the junction, since that is the regime assumed in the activation-energy study. Otherwise, a too strong dependence of the SBH with the metal-semiconductor bias, even leading to unphysical negative values for the SBH~\cite{LiGrassiLi2018}. Further information within the tunneling regime can be obtained with the analysis by Mouafo {\it et al.}~\cite{MouafoGodelFroehlicher2016}.

We have fitted the $n$-doped structure in \Fref{fig:iv-temp-all}.b) to \Eref{eq:E0vsT}, obtaining $E_{00} = 60.72$~meV and $\eta = 1.33$. So, for this case we are in the FE regime, with a small temperature assistance. The $p$-doped case is more complicated. We see in \Fref{fig:iv-temp-all}.c) how, specially at low temperatures, the curves present two regions with separate temperature dependence; at small bias the $T$ dependence is weak, becoming stronger at bias~$\gtrsim 0.25$~V. \Fref{fig:iv-temp-all}.d) shows $E_{0}$ extracted from a fit in the forward bias [0.3, 0.4]~V range, where we see that, at low $T$, we have some contribution of TFE current, overwhelmed by TE current at medium and large temperatures (the fitted parameters are $E_{00}=16.25$ and $\eta=2.04$).

\begin{figure}[t!]
	\centering
	\includegraphics[width=1.0\linewidth]{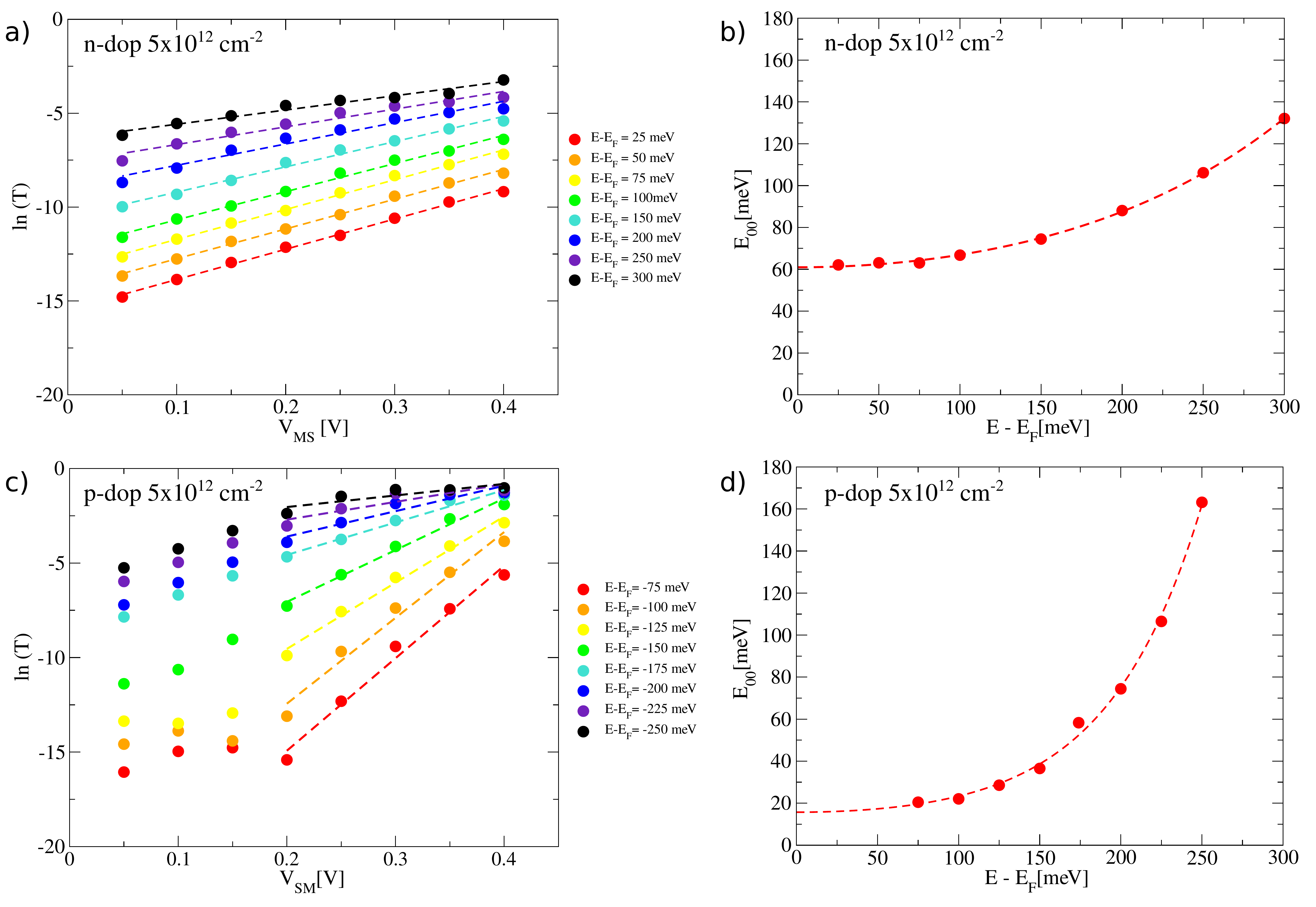}
	\caption{Logarithm of the transmission coeficients (T), at the indicated carrier incoming energies, as a function of the applied bias (dots), and linear regression with slope $S=1/E_{00}$ (dashed lines), for a) n-doped and c) p-doped structures. Plots b) and d) show the slope as a function of the excess energy $\delta E$ and its extrapolation to $\delta E = 0$.}
	\label{fig:GvsV}
\end{figure}

It is desireable to find $E_{00}$ by other means in order to check the consistency of the treatment and have further validation for the claimed transport regime.
In the case of a 3D structure and a parabolic barrier, $E_{00}$ can be easily evaluated from the metal-semiconductor (MS) junction parameters, finding that $E_{00}=q \sqrt{ N_D \hbar^2 / 4 \varepsilon_s m^\ast}$~\cite{PadovaniStratton1966}, where $N_D$ is the 3D dopant concentration, $\varepsilon_s$ is the semiconductor dielectric constant and $m^\ast$ is its effctive mass, under the parabolic dispersion assumption.
For the 2D MS junction, the barrier profile is no longer parabolic~\cite{electrostatic-pn}, difficulting the obtaining of an analytical expression. However, $E_{00}$ can be numerically estimated noting that the transmission coefficient for carriers coming at the Fermi level $E_F$ may be written as~\cite{Padovani1971}:
\begin{equation}
	 T(E_F) = \exp \left[ - q ( V_B - E_F - V) / E_{00} \right] ,
	\label{eq:T_EF}
\end{equation}
where $V_B$ is the Schottky barrier height in the semiconductor side and $V$ is the applied bias. We have carried out this approach, plotting the transmission coefficient as a function of the applied bias in \Fref{fig:GvsV} for the $n,p=5 \times 10^{12}$~cm$^{-2}$ doping at an incoming electron energy of $E_F + \delta E$~eV\footnote{The incoming energy with respect to the Fermi level must be increased (decreased) for electrons (holes) to ensure that the incoming carriers have allowed energies.} and fitting to \Eref{eq:T_EF}, for several $\delta E$. The obtained slopes have been extrapolated to $\delta E = 0$ with a cosh function, and from there we have obtained for the $n=5 \times 10^{12}$~cm$^{-2}$ case $E_{00}=61.0$~meV, in excellent agreement with the value obtained from the temperature analysis, thus corroborating the assignation to the FE regime. For the $p=5 \times 10^{12}$~cm$^{-2}$ case we have obtained $E_{00}=15.7$~meV, which, being quite smaller than $kT$ at room temperature, corroborates the assignment to the TE regime.

\subsection{Schottky barriers for 2H-1T' interfaces\label{ssec:sb}}

In experiments, Schottky barrier heights ($\Phi_{B}$) are normally obtained through the activation-energy method, which uses the thermionic emission equations to extract $\Phi_{B}$ from the $\ln (I/T^{\alpha})$ vs. $1/T$ plot. While this method can be applied to the intermediate $p$-doped case at high bias, see \Fref{fig:lnTvsT} in the SI, where we find an activation energy of 0.28~eV; it is not suitable for the intermediate $n$-doped and low bias $p$-doped cases because the doping concentrations we considered are high enough to observe transport governed by FE or TFE. Therefore, we studied the effect of the gate bias on the Schottky barriers from local density of states (LDOS) plots. Although penetration of the metallic states into the semiconductor gap renders the direct determination of the SBH difficult, we can use the macroscopic average of the Hartree portential, which traces quite closely the conduction and valence band profile~\cite{StradiMartinezBlom2016}, to extract the $n$-doped and $p$-doped barrier according to:  
\begin{eqnarray}
 \Phi_{B,n} &= q(V_{i} - V_{\it bulk}) + (E_{\it CBE} - E_{F}^{2H}) - (E_{F}^{1T'} - E_{F}^{2H})  \label{eq:barrier_n} \\
 \Phi_{B,p} &= q(V_{\it bulk} - V_{i}) + (E_{F}^{2H} - E_{\it VBE}) - (E_{F}^{2H} - E_{F}^{1T'})  \label{eq:barrier_p} ,
\end{eqnarray}
where $V_{bulk}$ is the average Hartree potential at the semiconductor side far from the influence of the interface, $V_{i}$ is the average Hartree potential at the interface, and $E_{CBE}$ and $E_{VBE}$ are the conduction band and valence band edges, respectively, extracted from the LDOS. Finally, in order to take into account the case of an applied bias across the junction, we define $ E_{F}^{1T'}$  and $ E_{F}^{2H}$ as the respective Fermi levels.

\begin{figure}[t!]
	\centering
	\includegraphics[width=0.5\linewidth]{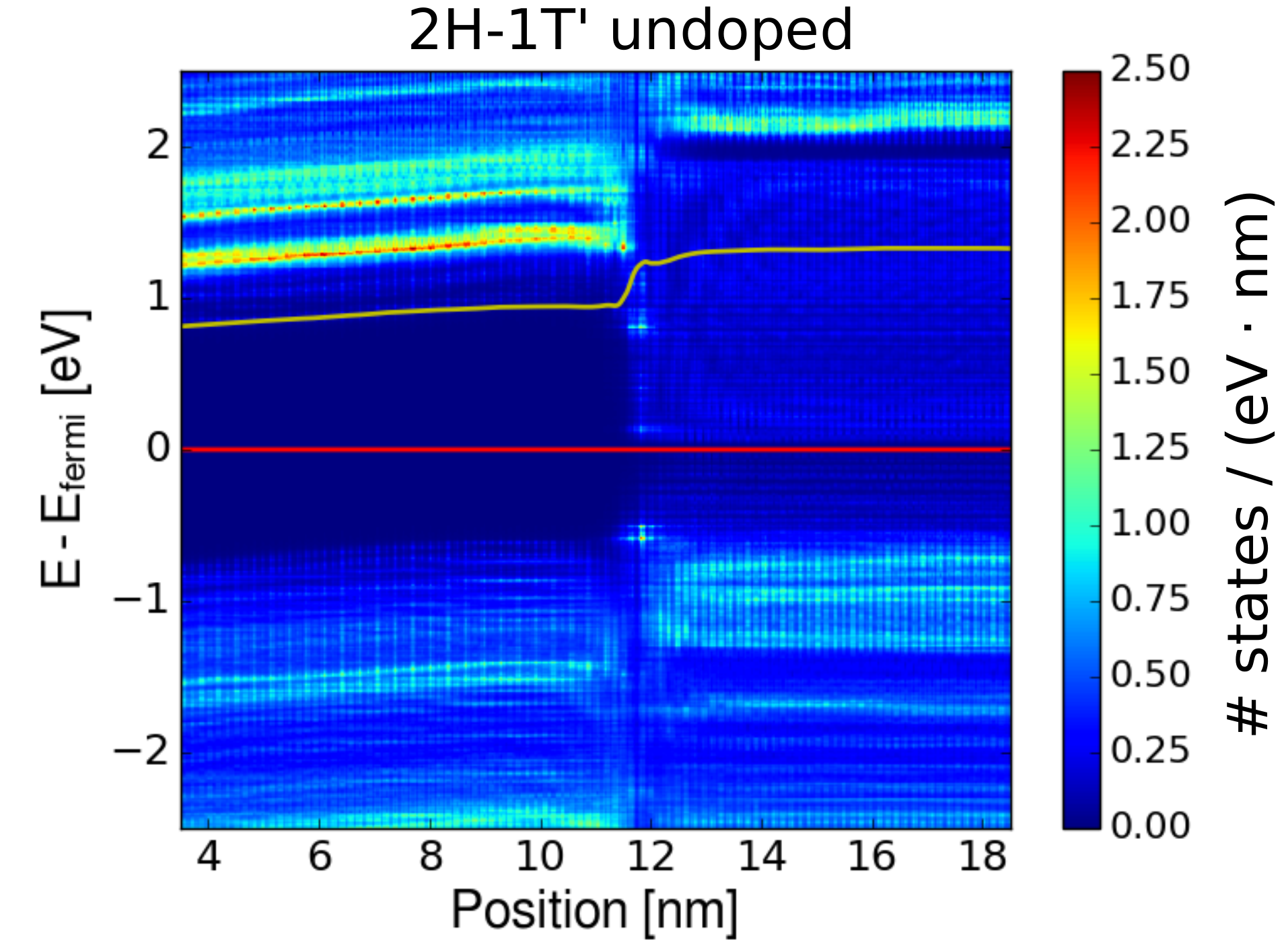}
	\caption{Energy-resolved local density of states (LDOS) of undoped \mos2 2H-1T' edge contacts with ac interface. The red and yellow lines denote the Fermi level and the average Hartree potential, respectively.}
	\label{fig:ldos-pristine}
\end{figure}

\begin{figure}[t!]
	\centering
	\includegraphics[width=1.0\linewidth]{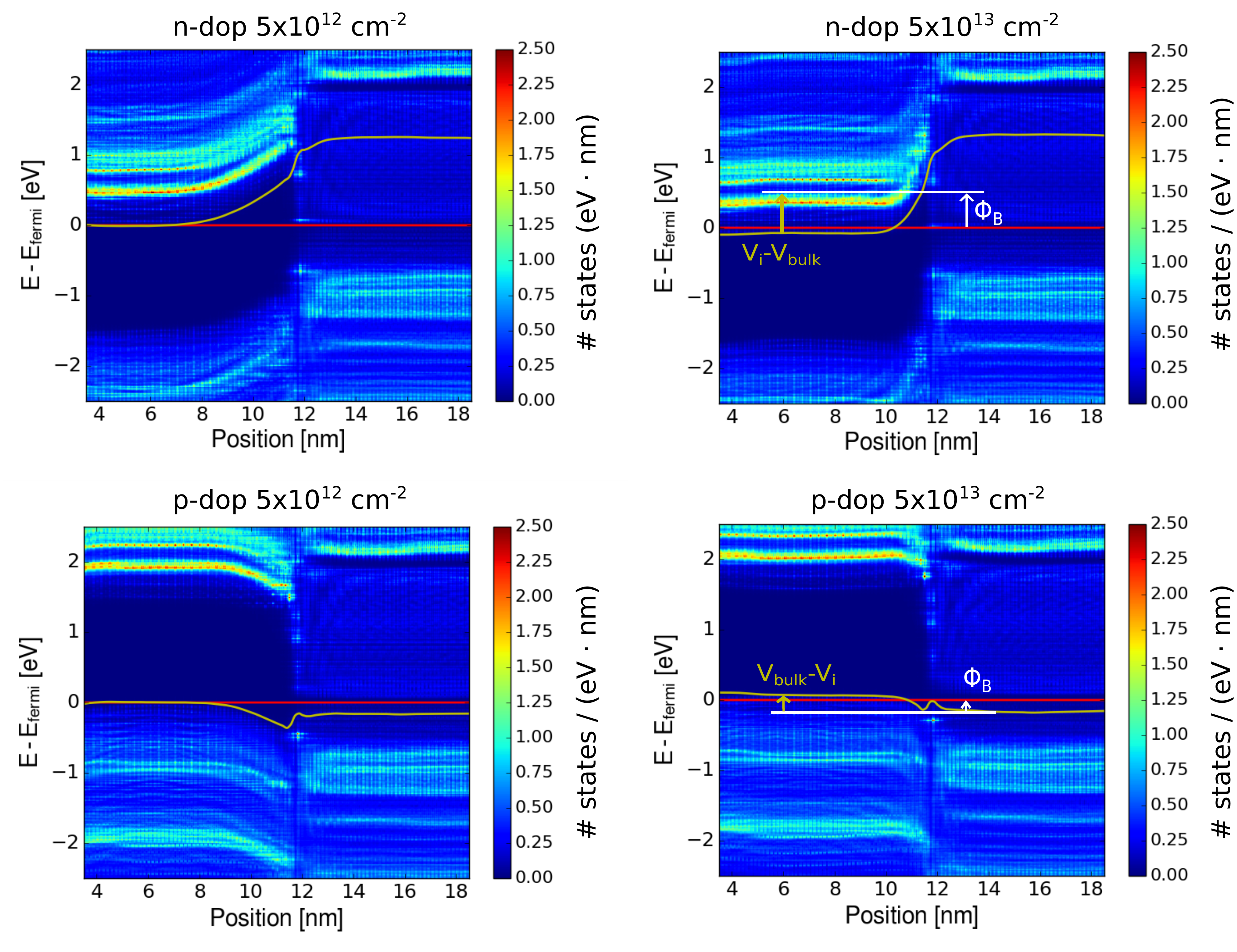}
	\caption{LDOS of armchair 2H-1T' structures with different $p$ and $n$ doping concentrations. See \Fref{fig:ldos-pristine} for meaning of lines.	}
	\label{fig:ldos-ac}
\end{figure} 
\begin{table}[t!]
	\centering
	{\renewcommand{\arraystretch}{1.3}
		\caption{Schottky barriers $\Phi_{B}$ and depletion layer widths (W) for positive and negative doping extracted from LDOS plots shown in Figures \ref{fig:ldos-pristine} and \ref{fig:ldos-ac}. \label{table:barriers-ac}} 
		\begin{tabular}{ c c c c c}
			\hline
			\multirow{2}{*}{$\delta q$ [cm $^{-2}$]} \ & \multicolumn{2}{c}{n-doped} & \multicolumn{2}{c}{p-doped} \\
			\cline{2-5}
			&  $\Phi_{B,n}$ [eV]  & W [nm] & $\Phi_{B,p}$ [eV] & W [nm] \\
			\hline
			0 \ & \ 0.97 \ & \ $>$8.6 \ & \ 0.58 \ & \ $>$8.6 \\
			5$\times$10$^{12}$ \ & \ 0.78 \ & \ 4.5 \ & \ 0.39 \ & 2.9 \\
			5$\times$10$^{13}$ \ & \ 0.60 \ & \ 1.3 \ & \ 0.04 \ & 0.9 \\
			\hline
		\end{tabular}
	}
\end{table}

The LDOS plot for the undoped armchair 2H-1T' interface at zero bias is shown in \Fref{fig:ldos-pristine}. From the plot, it can be seen that $\Phi_{B}$ for the hole injection is lower than for electrons, the values being 0.60~eV and 0.97~eV, respectively. This result is in good agreement with other theoretical values reported in literature, where the height of the $p$ and $n$-barriers were found to be 0.71 eV and 0.96~eV, respectively~\cite{saha}.

The LDOS plots for $n$-doped and $p$-doped armchair (zigzag) structures at zero bias are shown in \Fref{fig:ldos-ac} (\Fref{fig:ldos-zz-all}, in SI).
The barrier heights ($\Phi_{B}$) and the depletion layer width ($W$) extracted from Figures \ref{fig:ldos-pristine} and \ref{fig:ldos-ac} (\Fref{fig:ldos-zz-all}) are presented in Table~\ref{table:barriers-ac} (Table~\ref{table:barriers-zz}).
Of course, the depletion layer width is reduced as the doping level increases, leading to the ohmic-like behavior of the contact seen in \Fref{fig:iv}.
We also observe that the Schottky barrier decreases when the electrostatic doping increases, and it slightly varies for ac and zz interfaces. This evidence indicates that Schottky-Mott Rule does not apply in these highly doped 2H-1T' semiconductor-metal juctions, as expected for non-ideal systems~\cite{SB-Review}. A similar behavior was also observed in DFT studies of a Ag/Si 3D junction, where increase of the semiconductor doping level led to a reduction of the SBH~\cite{StradiMartinezBlom2016} by similar amounts. Note that this dependence of the SBH on the doping level is qualitatively different from that reported in graphene-silicon contacts~\cite{YangHeoPark2012}, because there the variation in electrostatic doping was applied to the graphene ``metallic'' component of the junction, thus changing the metal workfunction.
Additionally, our results point to the possibility that the 2H-1T' junction is free from Fermi level pinning, a possibility already hinted at by Katagiri {\it et al.}~\cite{Katagiri2016}, since the Fermi level position at the interface spans most of the semiconductor gap region (cf.\ the two 5$\times 10^{13}$~cm$^{-3}$ plots in \Fref{fig:ldos-ac} for the two extreme cases). This is opposite to junctions with 3D metals, where pinning of the Fermi level was found~\cite{GuoLiuRobertson2015, GongColombo2014, KimMoon2017}. This might open up the possibility of ambipolar injection into \mos2, similarly to what has been observed in MoTe$_2$ under weak Fermi level pinning conditions~\cite{NakaharaiYamamotoUeno2016}.

\begin{table}[t!]
	\centering
	{\renewcommand{\arraystretch}{1.3}
		\caption{Schottky barriers $\Phi_{B}$ for structures with different doping concentrations under finite bias, extracted from the LDOS plots shown in \Fref{fig:ldos-ac-ndop-voltage} and \Fref{fig:ldos-ac-pdop-voltage}. \label{table:barriers-ac-}} 
		\begin{tabular}{ c c c c c}
			\hline
			$\mu_{1T}$ \ & \  n-dop 5$\times$10$^{12}$ \ & \ n-dop 5$\times$10$^{13}$ \ & \ p-dop 5$\times$10$^{12}$ \ & \ p-dop 5$\times$10$^{13}$ \\
			\hline
			-0.4 eV \ & \ 0.85 \ & \ $-$  \ & \ 0.32 \ & \ $-$  \\
			-0.2 eV \ & \ 0.82 \ & \ 0.64 \ & \ 0.35 \ & \ 0.03 \\
			 0.0 eV \ & \ 0.79 \ & \ 0.58 \ & \ 0.39 \ & \ 0.15 \\
			 0.2 eV \ & \ 0.77 \ & \ 0.53 \ & \ 0.44 \ & \ 0.27 \\
			 0.4 eV \ & \ 0.74 \ & \ $-$  \ & \ 0.49 \ & \ $-$  \\
			\hline
		\end{tabular}
	}
\end{table} 
\begin{figure}[t!]
	\centering
	\includegraphics[width=1.0\linewidth]{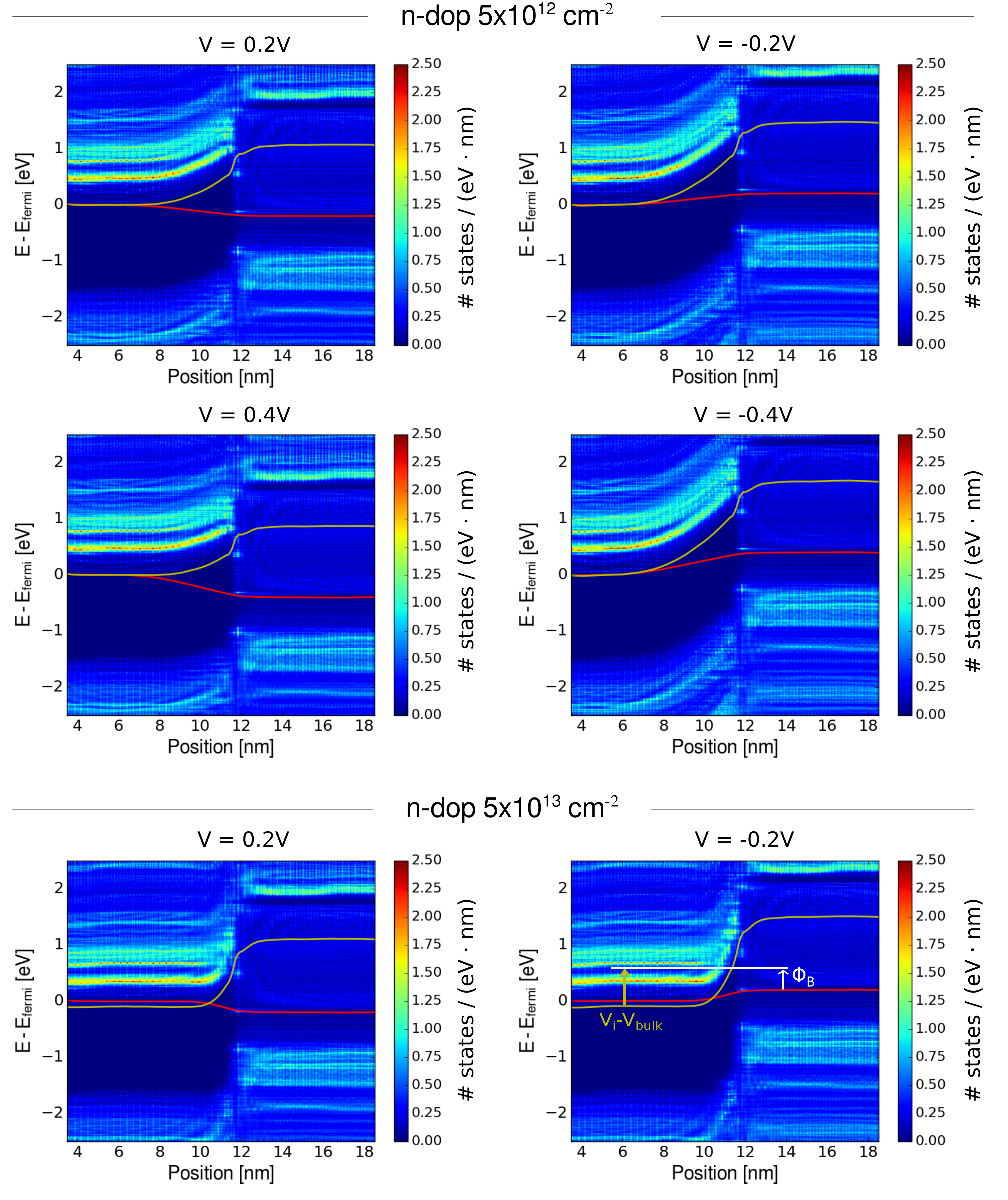}
	\caption{LDOS of n-doped 2H-1T' armchair structures at different applied voltage. The colorbar indicates the density of states. The red line is the difference between the averaged potential at finite bias and zero bias. The yellow line is average potential of the structure with the applied voltage. Positive (negative) voltage indicates forward (reverse) bias.}
	\label{fig:ldos-ac-ndop-voltage}
\end{figure}
\begin{figure}[t!]
	\centering
	\includegraphics[width=1.0\linewidth]{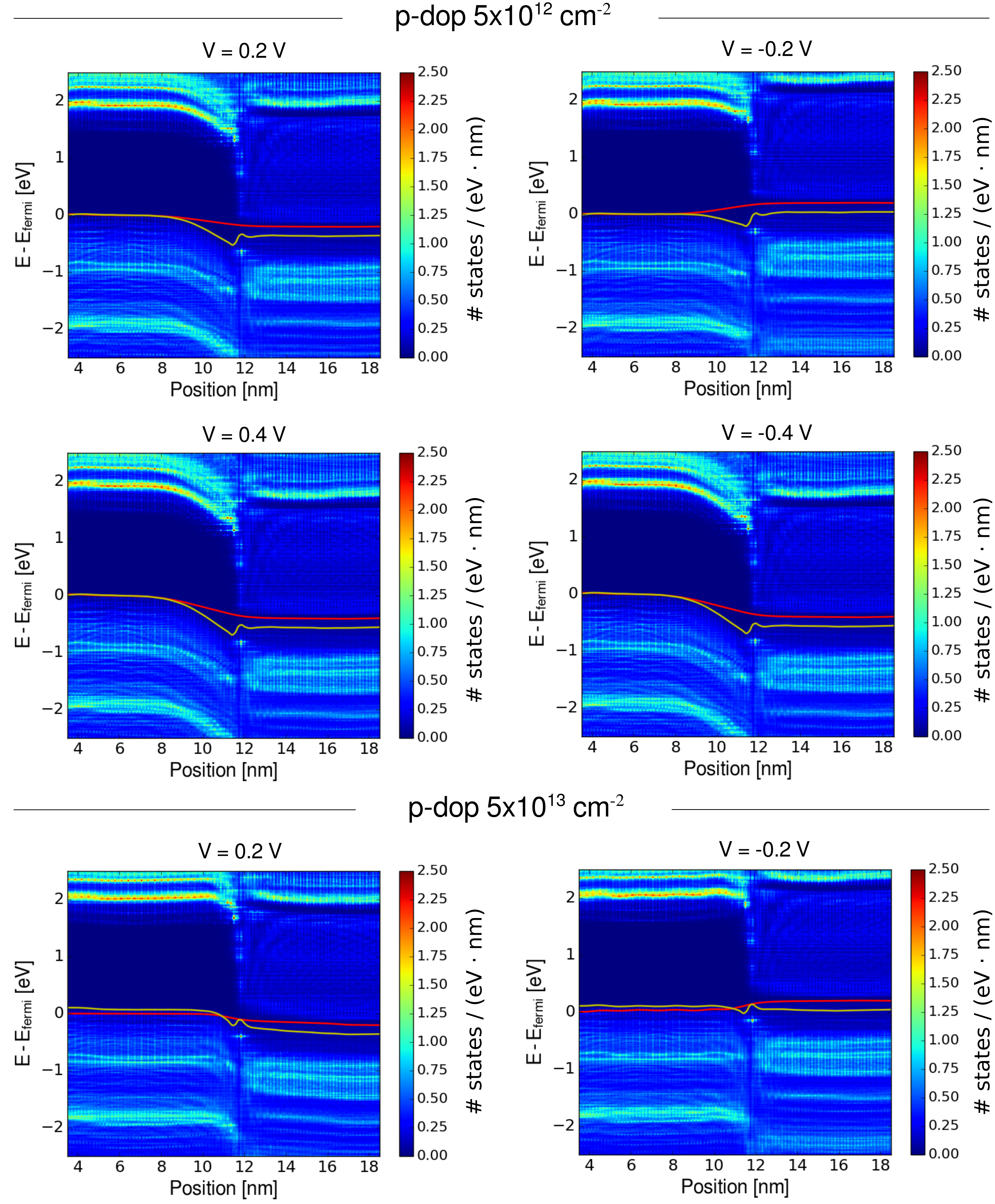}
	\caption{LDOS of 2H-1T' of p-doped armchair structures at different applied voltage. The colorbar indicates the density of states. The red line is the difference between the averaged potential at finite bias and zero bias. The yellow line is average potential of the structure with the applied voltage. Negative (positive) voltage indicates forward (reverse) bias.}
	\label{fig:ldos-ac-pdop-voltage}
\end{figure}

To understand the effect of the applied voltage on the band alignment of the 2H-1T' interfaces, we also represented the LDOS of the ac structures under finite bias, as can be seen in Figures~\ref{fig:ldos-ac-ndop-voltage} and \ref{fig:ldos-ac-pdop-voltage} for $n$-doped and $p$-doped structures, respectively. The $\Phi_{B}$ calculated according to Eqs. (\ref{eq:barrier_n}) and (\ref{eq:barrier_p}) are presented in Table~\ref{table:barriers-ac} for different biases.

In all cases we observe $\Phi_{B}$ slightly reduces (increases) when the 2H-1T' junction is reverse (forward) biased, again by amounts similar to what was observed in the Ag/Si junction~\cite{StradiMartinezBlom2016}, where this variation was attributed to the effect of image forces~\cite{Sze2007}.
For the case of intermediate n(p)-doping concentrations, 5$\times$10$^{12}$~cm$^{-2}$, when the 2H side is forward biased with respect to the 1T' phase, V$>$0 (V$<$0) in Figures \ref{fig:ldos-ac-ndop-voltage} and \ref{fig:ldos-ac-pdop-voltage}, the 2H band edge raises (lowers) and the effective barrier is reduced, increasing the current. 
When the junction is reverse biased the band bending increases, and so does the depletion layer; as result, only a small leakage current flows. 
On the other hand, when the transport is ohmic (high doping concentrations 5$\times$10$^{13}$~cm$^{-2}$), the band bending is almost imperceptible and the current flows independently of the bias polarization in both types of doping.

 
\subsection{Contact resistance for 2H-1T' interfaces\label{sec:rc}} 

\begin{figure}[t!]
	\centering
	\includegraphics[width=1.0\linewidth]{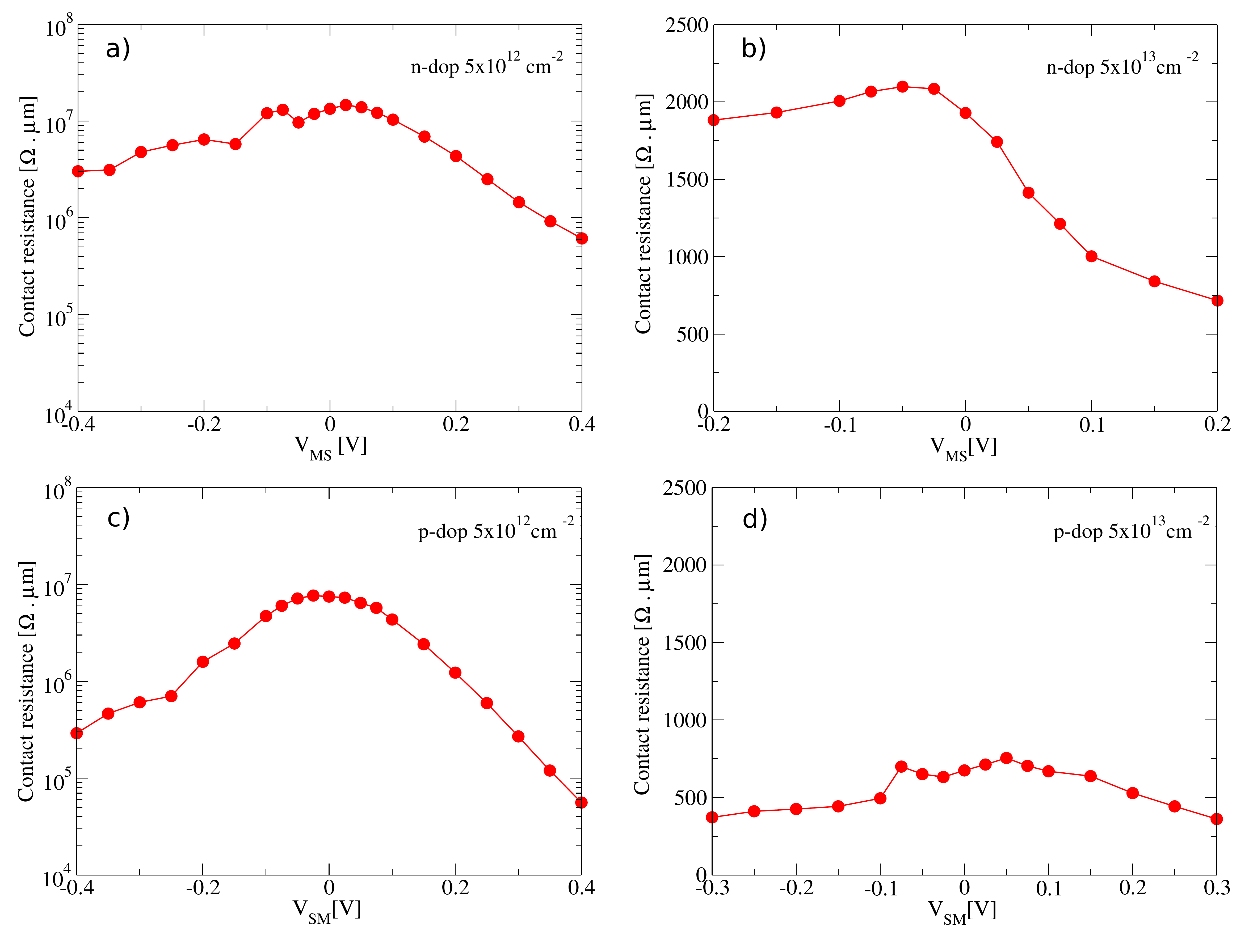}
	\caption{Contact resistance as a function of the voltage applied across the junction for structures with different electrostatic doping at the 2H phase.}
	\label{fig:rc-large-all}
\end{figure}

We calculated the large signal contact resistance ($R_c$), relevant for digital applications, of the 2H-1T' interfaces ($R_{2H-1T'}$) at 300~K using Eq.~(\ref{eq:rc_large}) in \sref{ssec:resistance}:

\begin{eqnarray}
	R_{2H-1T'} = \frac{V_{2H-1T'}}{I} \ - \ 
	\frac{1}{2} \ \frac{V_{2H-2H}}{I} \ - \ 
	\frac{1}{2} \ \frac{V_{1T'-1T'}}{I}
	\label{eq:rc_large_mos2} \ ,
\end{eqnarray}
where the voltages $V_{2H-1T'}$, $V_{2H-2H}$ and $V_{1T'-1T'}$ have been extracted from the I--V curves shown in \Fref{fig:iv-ref-all}, at the same current value $I$.
The resulting contact resistances for different electrostatic dopings as a function of $V_{MS/SM}$ are shown in \Fref{fig:rc-large-all}. We also show a comparison of the large- and small-signal contact resistances, in \Fref{fig:iv-ref-all}, finding a qualitatively similar behavior.

As expected from the I--V curves in \sref{ssec:transport}, positive doping yields lower contact resistance than negative doping by a factor of $\sim \!\! 10$ ($\sim \!\! 4$) in the intermediate (high) doping case.
We also observe that in the intermediate-doping/Schottky case the contact resistance has an exponential dependence with $V_{MS}$, very clear at forward bias. Having established in Sec.~\ref{ssec:transport} that the junction operates in the FE (TE) regime for the $n$ ($p$) case, the exponential decrease of $R_c$ at forward bias is due to the narrowing and lowering of the barrier seen from the semiconducting side ($V_{bi}$, see \Fref{fig:ldos-ac}). 
In the reverse bias, the lowering of $R_{2H-1T'}$ is due to the narrowing of the tunnel barrier for a fixed energy of the carrier coming from the metal side. 
In the high-doping/ohmic case, a linear drop in $R_{2H-1T'}$ is observed at forward bias. This is due to the (linearly) increased transmission as bias is raised.

Houssa {\it et al.}\ also performed calculations of contact resistances for 2H-1T---as opposed to 1T'---lateral heterojunctions but without including electrostatic doping. They obtained $R_{c}$ values on the order of 40 and 30~k${\rm \Omega \! \cdot \! \mu m}$ for armchair and zigzag interfaces, respectively, using the transfer length method~\cite{Houssa2019}. These values are much lower than our results for intermediate doping; however, the differences may be related to the procedure of computation of the $R_{c}$. They used the transfer length method, with semiconductor lengths up to 4.8~nm. As we can see in \Fref{fig:macroaverage}, the depletion width for the undoped case is $>8$~nm, meaning that they treated the fully depleted case, with some amount of indirect doping from the metal contacts.

We also compared our results to experimental measurements of \mos2-based FETs with metal-semiconductor junctions. Nourbakhsh {\it et al.}\ reported $R_{c}$ values of 1~k${\rm \Omega \! \cdot \! \mu m}$  at gate voltages of 3.5~V, i.e.\ for sufficiently high negative charge induced in the channel~\cite{sub-10nm-fet}. This result is in agreement with our calculations for negative 5$\times$10$^{13}$~cm$^{-2}$ doping concentration at forward bias. 

On the other hand, Kappera {\it et al.} reported $R_{c}$ values of 0.24~k${\rm \Omega \! \cdot \! \mu m}$ at zero gate voltage~\cite{kappera}. This extreme low value, obtained without inducing any charge in the channel, has been explained by considering the functionalization of 1T' phase with chemical dopants (H, Li, or H$_{2}$O), present during the local transformation of semiconducting 2H phase into metallic 1T' phase \cite{Houssa2019}.


\section{Summary \label{sec:conclusions}}

We used non-equilibrium Green's function formalism to perform transport calculations at finite voltage of armchair and zigzag 2H-1T' interfaces, where an electrostatic doping was added to the semiconducting 2H phase to emulate the gate voltage on the channel. 
It was observed that armchair interfaces provide better conductance as result of the anisotropic transport behavior of the 1T' phase. 
Besides, from the I--V analysis it was found that (i) electronic transport follows ohmic and Schottky regimes in highly and intermediate doped structures, respectively, (ii) for the Schottky case, the transmission occurs by tunneling in the intermediate $n$-doped structure and by thermionic emission in the $p$-doped structure.

The Schottky barrier heights of structures under different doping concentration and finite bias were obtained through their LDOS, observing that the barrier heights in 2H-1T' structures are sensitive to the applied voltage, both at the gate and at the semiconductor, and there was no indication of the presence of Fermi level pinning.
We also computed the contact resistance for different dopant types and concentrations as a function of the voltage applied across the junction, finding a lower 2H-1T' contact resistance for the $p$-doped 2H phase.

Finally, we have also pointed out a method that can be used to experimentally identify the emission regime (i.e.\ tunnel or thermoionic), prior to a possible use of the activation-energy method.

\ack
We acknowledge financial support by Spain's Ministerio de Econom\'ia, Industria y Competitividad under grant TEC2015-67462-C2-1-R (MINECO/FEDER), the Ministerio de Ciencia, Innovación y Universidades under Grant No. RTI2018-097876-B-C21 (MCIU/AEI/FEDER, UE), and the EU Horizon2020 research and innovation program under grants No. GrapheneCore2 785219 and GrapheneCore3 881603.

\vskip0.5cm

\bibliographystyle{iopart-num}
\bibliography{file}

\makeatletter\@input{xx.tex}\makeatother
\end{document}


\title[Schottky barriers, emission regimes and contact resistances in 2H-1T' \mos2 junctions]
      {Schottky barriers, emission regimes and contact resistances in 2H-1T' \mos2 lateral metal-semiconductor junctions from first-principles}

\author{M. Laura Urquiza}
\address{Departament d'Enginyeria Electr\`onica, Universitat Aut\`onoma de Barcelona, 08193 Bellaterra, Spain}
\ead{Laura.Urquiza@uab.es}

\author{Xavier Cartoix\`a}
\address{Departament d'Enginyeria Electr\`onica, Universitat Aut\`onoma de Barcelona, 08193 Bellaterra, Spain}
\ead{Xavier.Cartoixa@uab.es}
\maketitle

\section{Transport properties of ac and zz interfaces\label{ssec:SI-transport}}

\begin{figure}[t!]
	\centering
	\includegraphics[width=1.0\linewidth]{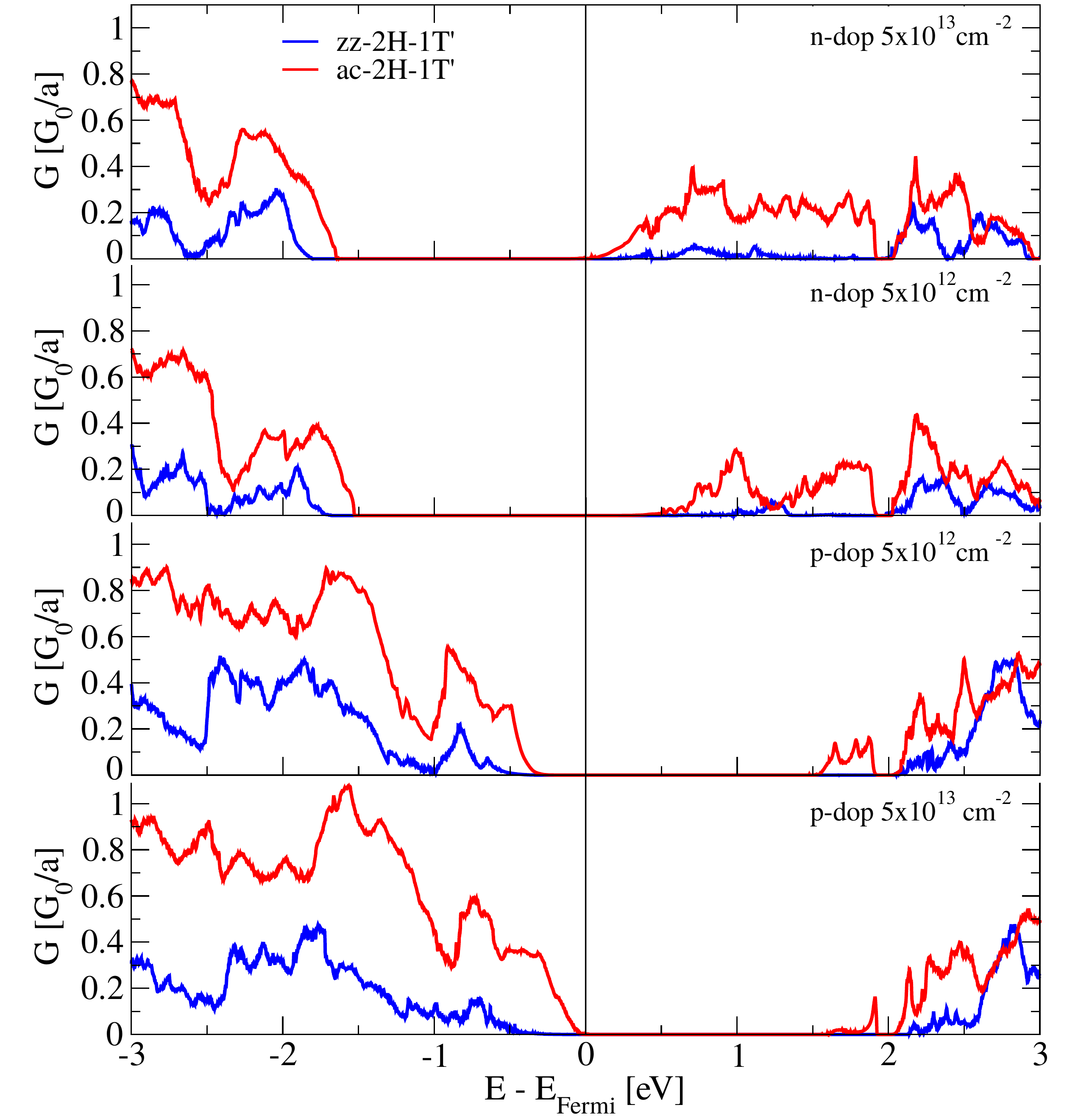}
	\caption{Specific conductance, in units of $G_{0}$ over the lattice parameter ($a_{0}$), at zero bias for armchair and zigzag 2H-1T' interfaces with different doping concentration.}
	\label{fig:avtrans-ac-zz}
\end{figure}

\begin{figure}[t!]
	\centering
	\includegraphics[width=0.8\linewidth]{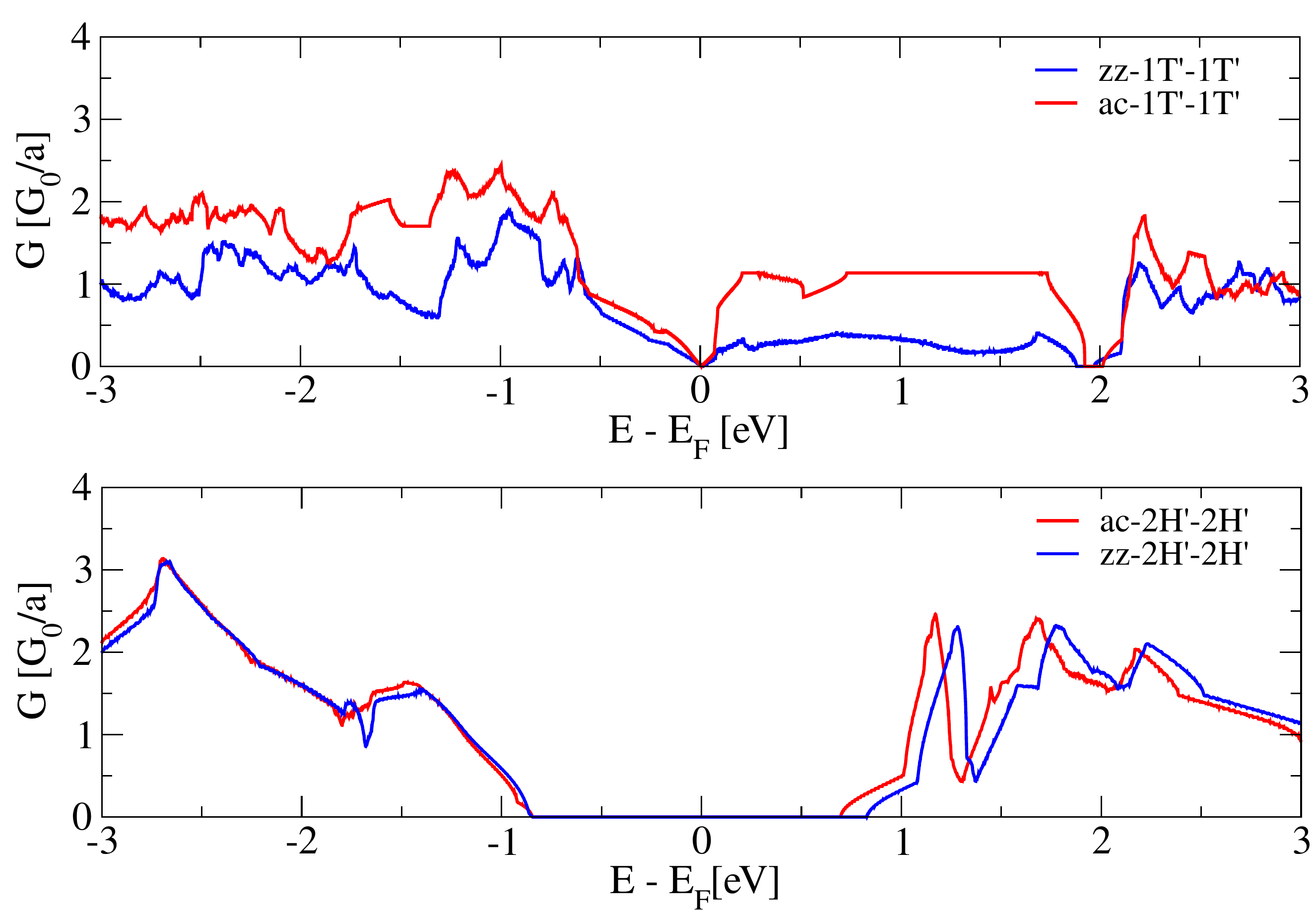}
	\caption{Conductance of the 2H and 1T' phases at Vg = 0. The orientation in the legend corresponds to the interface; i.e. zz-2H-2H means a zigzag interface between two 2H layers, transport thus being along the armchair direction.}
	\label{fig:avtrans-ref}
\end{figure}

We compared the transport properties of zigzag (zz) and armchair (ac) structures in order to predict which is the most promising configuration for transport. In \Fref{fig:avtrans-ac-zz} we show the specific conductance as a function of the energy of the incoming particle, for armchair and zigzag 2H-1T' interfaces.
The results show that the ac interface has an enhanced transmission for the injection of both holes and electrons. The difference in the transport across the ac and zz interfaces comes essentially from the asymmetrical behaviour of the 1T' phase, as can be seen in \Fref{fig:avtrans-ref}, which has higher conductance when transport direction is perpendicular to the ac interface. 
As for the 2H phase, it can be noticed that there is a small difference in the bandgap of the zz and ac structures, which is caused by the different strains induced along the armchair and zigzag directions to lattice match with the 1T' phase.


\section{Schottky barriers for 2H-1T interfaces\label{ssec:SI-sb}}	

\begin{figure}[t!]
   \centering
   \includegraphics[width=1.0\linewidth]{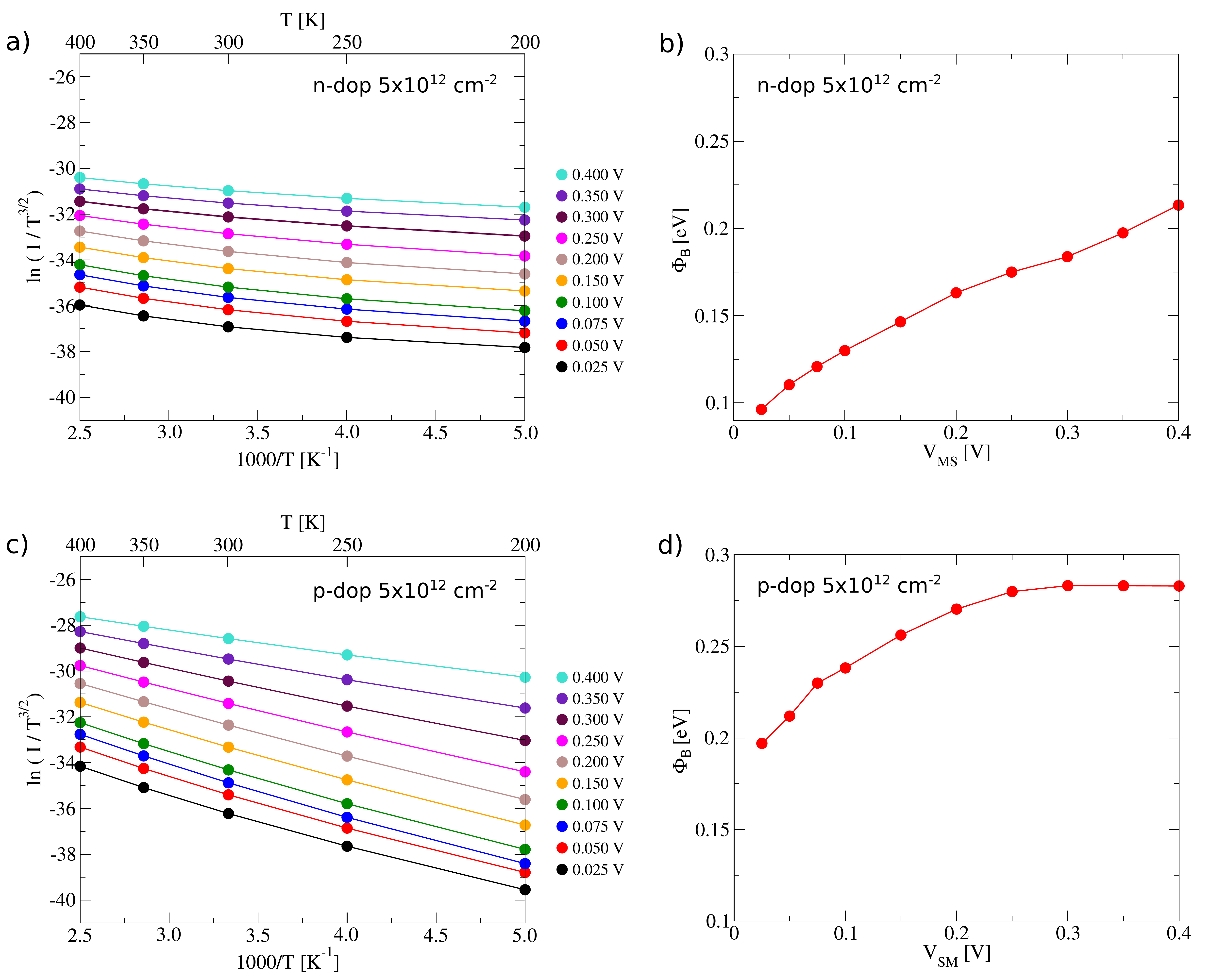}
   \caption{ Arrhenius plots of $\ln(I_{ds}/T^{3/2})$ vs. 1/T at different gate voltages for a) n-doped and c) p-doped suctures, and Schottky barrier height ($\Phi_{B}$) for b) n-doped and d) p-doped structures extracted from the activation energy method at different bias.}
   \label{fig:lnTvsT}
\end{figure}

\begin{figure}[t!]
	\centering
	\includegraphics[width=1.0\linewidth]{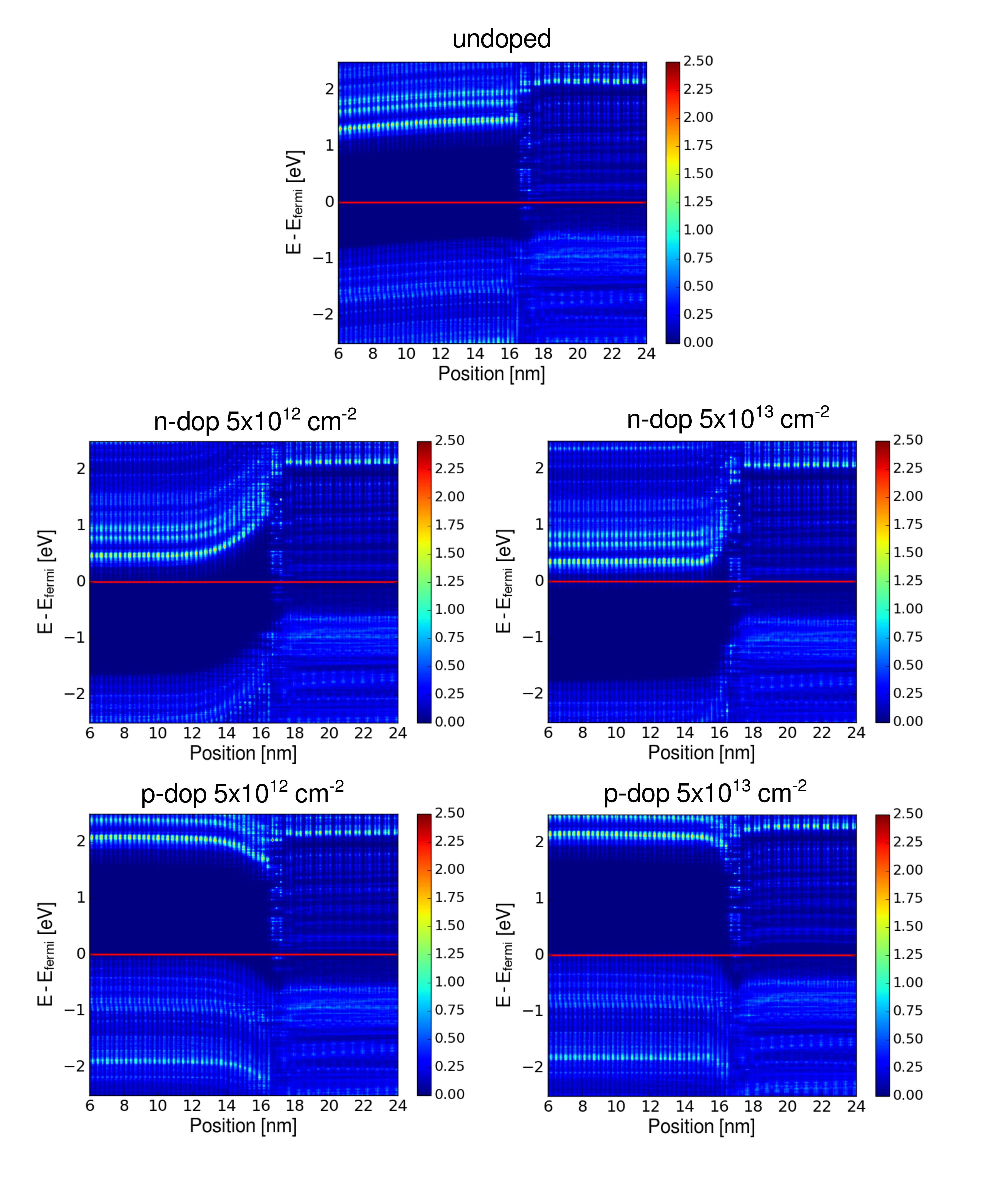}
	\caption{Energy-resolved local density of states (LDOS) of zigzag 2H-1T' edge contacts with different electrostatic doping at the semiconducting phase.}
	\label{fig:ldos-zz-all}
\end{figure}

We applied the activation energy method in \Fref{fig:lnTvsT} to illustrate its failure to provide a definite value for the Schottky barrier height in the intermediate doping structures.

We also compared the effect of the electrostatic doping on the Schottky barrier by plotting the energy-resolved local density of states (LDOS) for different positive and negative doping concentrations, as can be seen in \Fref{fig:ldos-zz-all}. From these plots, we extracted the Schottky barrier heigh and depletion layer width, the values are reported in Table \ref{table:barriers-zz}.   

\begin{table}[t!]
	\centering
	{\renewcommand{\arraystretch}{1.3}
		\caption{Schottky barriers $\Phi$ and depletion layer widths (W) for 
			positive and negative gating extracted from LDOS plots shown 
			in \Fref{fig:ldos-zz-all}. \label{table:barriers-zz}} 
		\begin{tabular}{ c c c c c}
			\hline
			\multirow{2}{*}{$\delta q$ [cm $^{-2}$]} \ & \multicolumn{2}{c}{n-doped} & \multicolumn{2}{c}{p-doped} \\
			\cline{2-5}
			&  $\Phi$ [eV]  & W [\AA] & $\Phi$ [eV] & W [\AA] \\
			\hline
			0 \ & \ 1.0 \ & \ $>$8.6 \ & \ 0.7 \ & \ $>$8.6 \\
			5$\times$10$^{12}$ \ & \ 0.8 \ & \ 4.1 \ & \ 0.4 \ & 3.2 \\
			5$\times$10$^{13}$ \ & \ 0.2 \ & \ 1.1 \ & \ 0.2 \ & 1.0 \\
			\hline
		\end{tabular}
	}
\end{table}

\section{Contact resistance for 2H-1T' interfaces\label{sec:SI-rc}} 

\begin{figure}[t!]
	\centering
	\includegraphics[width=1.0\linewidth]{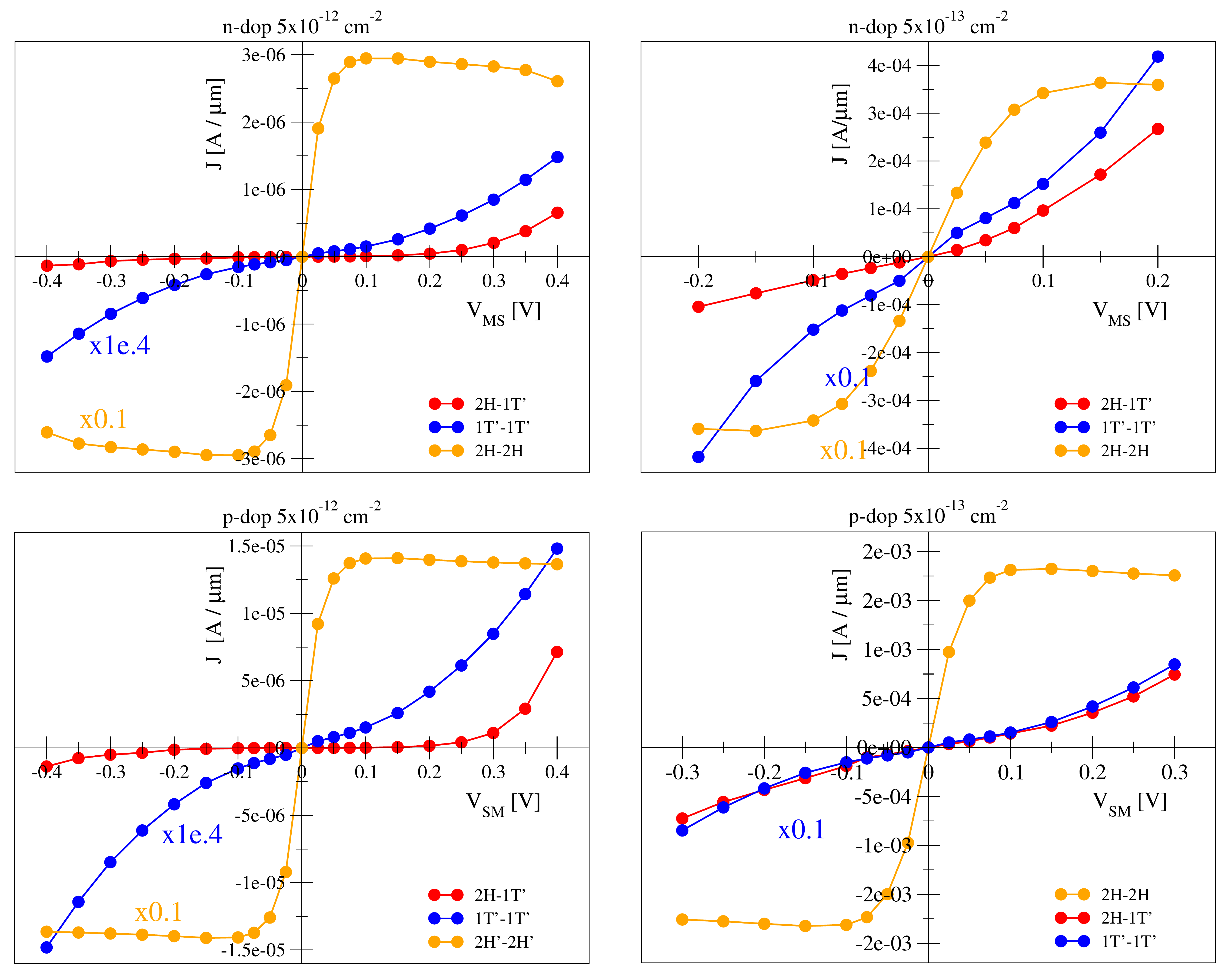}
	\caption{Current vs. voltage curve for the 2H-1T' device and reference devices (2H-2H and 1T'-1T') with different doping concentration.}
	\label{fig:iv-ref-all}
\end{figure}

\begin{figure}[t!]
	\centering
	\includegraphics[width=1.0\linewidth]{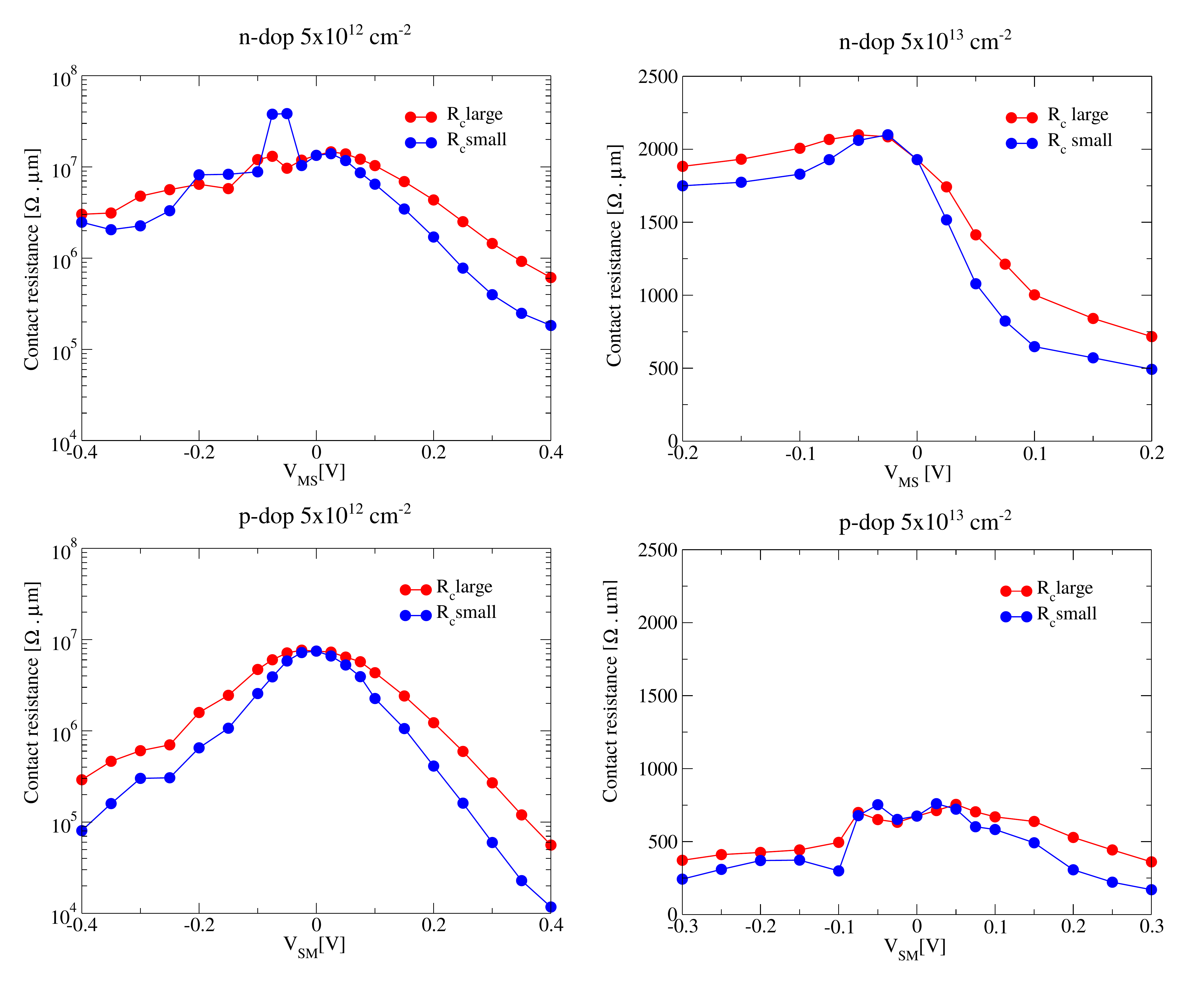}
	\caption{Small and large signal contact resistances as a function of V$_{ds}$ for 2H'-1T' contacts with different doping concentration.}
	\label{fig:rc-small-all}
\end{figure}

We calculated the contact resistance for small signal according to:
\begin{eqnarray}
	R_{2H-1T'} ^{small} \ \Bigg|_{I_{0}} = \
	\frac{\partial V_{tot}}{\partial I} \ \Bigg|_{I_{0}} - \ 
	\frac{1}{2} \ \frac{\partial V_{2H-2H}}{\partial I} \ \Bigg|_{I_{0}} - \ 
	\frac{1}{2} \ \frac{\partial V_{1T'-1T'}}{\partial I} \ \Bigg|_{I_{0}}
	\label{eq:rc_small_mos2}
\end{eqnarray}
where the voltages $V_{tot}$, $V_{2H-2H}$ and $V_{1T'-1T'}$ have been extracted from the I-V curves shown in \Fref{fig:iv-ref-all} at the same current value. 
The results are compared with large-signal contact resistance in \Fref{fig:rc-small-all}, for different doping concentrations. 

